\documentclass[a4,twoside,3pt,times,12pt]{elsarticle}
\biboptions{numbers,sort&compress}
\usepackage{graphicx,enumitem} 
\usepackage{geometry}
\usepackage{amssymb}
\usepackage{amsmath}
\usepackage{subfigure}
\usepackage{epstopdf}
\usepackage{bibentry}
\usepackage{caption}
\usepackage{color,xcolor}
\usepackage{mathrsfs}
\usepackage{multirow}
\usepackage{makecell}
\usepackage{hhline}
\usepackage{breqn}
\usepackage{fancyhdr}
\usepackage{titlesec}
\usepackage{lineno,hyperref}
\titleformat{\section}
  {\Large\bfseries}{\thesection}{0.5em}{}
\titleformat{\subsection}
  {\large\bfseries}{\thesubsection}{0.5em}{} 
\titleformat{\subsubsection}
  {\normalfont\bfseries}{\thesubsubsection}{0.5em}{} 
\titlespacing*{\section}{0pt}{1.5ex plus 0.5ex minus .2ex}{1.3ex plus .2ex}
\titlespacing*{\subsection}{0pt}{1.2ex plus 0.5ex minus .2ex}{1ex plus .2ex}
\DeclareMathOperator{\sech}{sech}
\numberwithin{equation}{section}

\newtheorem{prop}{Proposition}[section]

\newtheorem{rk}{Remark}[section]

\allowdisplaybreaks[4]

\def\e{{\rm e}}

\def\e{{\rm e}}

\hypersetup{
    colorlinks=true,      
    linkcolor=blue,       
    filecolor=red!70!blue,       
    urlcolor=red,         
    citecolor=green!70!black,      
}

\journal{Studies in Applied Mathematics}
\begin{document}
\begin{frontmatter}
\title{{\bfseries The variable-length stem structures in three-soliton resonance of the Kadomtsev-Petviashvili II equation}\vspace{-0.5\baselineskip}}

\author[a]{Feng Yuan \corref{cor2}}
\author[b]{Jingsong He\corref{cor1}}
\author[c]{Yi Cheng}
\address[a]{College of Science, Nanjing University of Posts and Telecommunications, Nanjing, 210023, P. R. China}
\address[b]{Institute for Advanced Study, Shenzhen University, Shenzhen, 518060, P. R. China}
\address[c]{School of Mathematical Sciences, USTC, Hefei, Anhui 230026, P. R. China}

\cortext[cor1]{Corresponding author. hejingsong@szu.edu.cn}
\cortext[cor2]{Corresponding author. fengyuan@njupt.edu.cn}

\begin{abstract}
The stem structure is a localized feature that arises during high-order soliton interactions, connecting the vertices of two V-shaped waveforms. 
The interaction of resonant 3-solitons is accompanied by soliton reconnection phenomena, characterized by the disappearance and reconnection of stem structures.
This paper investigates variable-length stem structures in resonant 3-soliton solutions of the Kadomtsev-Petviashvili II (KPII) equation, focusing on both 2-resonant and 3-resonant 3-soliton cases. 
Depending on the phase shift tends to plus/minus infinity, different types of resonances are identified, including strong resonance, weak resonance, and mixed (strong-weak) resonance. 
We derive and analyze the asymptotic forms and explicit expressions for the soliton arm trajectories, velocities, as well as the endpoints, length, and amplitude of the stem structures. 
A detailed comparison is made between the similarities and differences of the stem structures in the 2-resonant and 3-resonant solitons. 
In addition, we provide a comprehensive and rigorous analysis of both the asymptotic behavior and the structural properties of the stems.

\end{abstract}
\end{frontmatter}
\vspace{-\baselineskip}

\section{Introduction}
Solitons, as one of the most fundamental and representative wave in nonlinear science, have garnered significant attention since the term was first observed by Russell \cite{russel} and named by Zabusky and Kruskal \cite{prl1965}. 
These localized wave packets arise from a delicate balance between nonlinearity and dispersion, exhibiting particle-like stability and elastic interaction properties \cite{book04,prl1965,cpam1974}. 
The most remarkable feature of a soliton lies in its ability to maintain its shape during propagation and to undergo collisions with other solitons without losing its identity \cite{book04,cpam1974,book03,interaction03}.
This behavior not only embodies profound mathematical elegance but also underpins its relevance across a broad spectrum of physical disciplines.

Solitons play essential roles in nonlinear optics, fluid dynamics, plasma physics, and Bose-Einstein condensates, among others \cite{book01,soliton01,soliton02,science2002,pnas2017}. 
As such, soliton models have emerged as indispensable theoretical tools for understanding key phenomena in complex nonlinear systems, particularly regarding energy localization and the formation and evolution of coherent structures. 
With the advancement of soliton theory, researchers gradually recognized the limitations inherent in classical (1+1)-dimensional integrable models, such as the Korteweg-de Vries (KdV) \cite{kdv1895,prl1967} and nonlinear Schr\"{o}dinger (NLS) equations \cite{nls1,nls2}, when applied to real-world phenomena. 
Many nonlinear wave processes in nature exhibit intrinsically higher-dimensional behavior, such as transverse diffraction in shallow water waves, multi-directional disturbances in plasmas, or transverse mode coupling in optics \cite{multi1,multi2,multi3,multi4}. 
These realities necessitate the development of higher-dimensional integrable models that can faithfully capture such effects while preserving mathematical tractability. 

In response to this need, B. B.  Kadomtsev and V. I.  Petviashvili introduced the Kadomtsev-Petviashvili (KP) equation--including KPI and KPII equation--in 1970 to describe weakly nonlinear, weakly dispersive long waves in quasi-two-dimensional media \cite{kp1970}. 
The Kadomtsev-Petviashvili II equation is given by:
\begin{equation}\label{kpeq}
(u_t+6uu_x+u_{xxx})_x+3u_{yy}=0.
\end{equation}
The KPII equation is not merely a straightforward extension of the KdV equation to accommodate transverse perturbations; 
it also retains a rich integrable structure, including a Lax pair, bilinear form, and an infinite hierarchy of conservation laws \cite{kp1982,kp2013,kp2014-1,kp2014-2,kp2015,book2}. 
These mathematical features allow for the systematic construction of multi-soliton and resonant solutions via analytical techniques such as the Hirota bilinear method, the Darboux transformation, and the inverse scattering method. 

Compared to the simple elastic interactions of one-dimensional solitons, soliton interactions under the KPII framework exhibit a far greater degree of richness and structural complexity \cite{kp1977,jpsj1980,jpsj1983}. 
In particular, when the phase of multiple line solitons satisfy specific conditions, the resulting solutions manifest as resonant solitons, a special class of nonlinear structures \cite{kp1977,jpsj1980,jpsj1983,kp01,kp02,kp03,kp04}. 
This phenomenon has been systematically investigated by Chakravarty, Kodama, and others, who employed geometric and combinatorial tools such as Weyl chambers and the Grassmannian manifold to classify line soliton interactions and reveal the conditions under which resonant patterns emerge \cite{kp2004,kp2008,kp2009,resonance2014}.

Under resonance conditions, the incoming solitons do not merely pass through each other elastically; 
instead, they undergo a process of fusion or fission, leading to the formation of stable, coherent wave networks in the interaction region \cite{resonance1980,resonance1993,resonance2010}. 
The mathematical foundation of this behavior lies in the $\tau$-function representation of the KPII solutions, where specific exponential terms constructively interfere to generate localized, stationary interference patterns \cite{resonance2010,resonance2022-1,resonance2022-2}. 
These structures do not dissipate over time but persist as spatially organized configurations, commonly taking the form of Y-type, X-type, and more complex network-type \cite{resonance2012,resonance2014}. 
Such patterns are the hallmarks of coherents resonant interactions among multiple solitons.

Although existing studies have systematically revealed the overall structure and evolution characteristics of resonant soliton networks in the KPII equation, a class of highly localized substructures
---referred to in this paper as stem structures---remains embedded within these complex configurations and has yet to be thoroughly explored in terms of their formation mechanisms, dynamical behavior, and localization properties \cite{kp1977,resonance2010}. 
A stem structure is a localized configuration that connects the two vertex tips of V-shaped solitons, and it appears in both resonant and quasi-resonant soliton interactions \cite{yuan2024,annv2025}. 
Previous work Refs.~\cite{yuan2024,annv2025} has conducted a systematic study of stem structures in the ANNV equation. 
And for the KPII equation, there have been several studies on multi-soliton solutions with stem-like structures. 
In Ref.\ \cite{jpsj1983}, the asymptotic form and graphical representation (e.g., Fig. 4) of quasi-resonant two-soliton solutions are presented, where the stem structure---referred to as a ``virtual soliton" in the paper---corresponds to the interactions between $u^{(1-2)}$ and $u^{(1+2)}$. 
For resonant three-soliton solutions, only certain types are discussed, and the asymptotic behavior is analyzed for the four outer arms, excluding the stem structure. 
Gino Biondini constructed lattice-like resonant soliton solutions with a hollow central region \cite{resonance1993,kp2003,kp2007}. 
These resonant solitons all contain localized structures, which can be interpreted as stem features (see, e.g., Ref.~\cite{kp2003} Figs.\ 5--8; Ref.~\cite{resonance1993} Figs. 4--5; Ref. \cite{kp2004} Figs. 4 and 7). 
The interactions of analytic multi-soliton solutions, including quasi-resonant two-soliton solutions, are investigated in \cite{kp02}. 
Further experimental and numerical studies on the interaction of multi-solitons with stem structures are reported in Refs.~\cite{kp01,kp04,kp2008}.
The characteristics of the stem profile---such as amplitude evolution and stability---were preliminarily explored using numerical solutions in Ref.~\cite{kp2011}. 
In 2012, Mark J. Ablowitz and Douglas E. Baldwin reported observations of quasi-resonant two-soliton water waves near low tide on two flat beaches approximately 2000 kilometers apart \cite{resonance2012}. 
However, their study provided only visual representations, without a systematic investigation of the stem's amplitude, speed, or a clear analytic description of this localized structure.
Y. Kodama conducted a systematic study of complex web-like patterns formed by soliton solutions of the KPII equation, including resonant solitons, from a geometric perspective. 
He developed a classification framework and examined its deep connections to the Grassmannian parametrization \cite{book2}.
Recently, the stem structures in quasi-resonant solitons have been comprehensively investigated by us \cite{kp2025}. 
However, analytical mathematical descriptions of stem structures within resonant solitons of this equation have not yet been specifically addressed in existing literatures. 
A detailed analysis of these structures can deepen our theoretical understanding of KPII equation solutions. 
It may also provide new insights and analytical methods for exploring how localized and extended patterns coexist in multidimensional nonlinear systems.

Motivated by these considerations and the significant difference from quasi-resonant case, 
the present work is devoted to a systematic investigation of the localized soliton structures in resonant solitons
---referred to as stem structures---that arise within the 3-soliton solutions of the KPII equation under specific resonant conditions, 
has the following form \cite{kp1976-1,kp1976-2}: 
\begin{flalign}\label{3f}
\begin{split}
&u=2(\ln f^{[3]})_{xx},\\
&f^{[3]}=1+\exp\xi_1+\exp\xi_2+\exp\xi_3+a_{12}\exp(\xi_1+\xi_2)+a_{13}\exp(\xi_1+\xi_3)\\
&+a_{23}\exp(\xi_2+\xi_3)+a_{12}a_{13}a_{23}\exp(\xi_1+\xi_2+\xi_3).
\end{split}
\end{flalign}
where,
\begin{flalign}\label{etakpw}
	\begin{split}
		\xi_j=k_jx+p_jy+\omega_j t+\xi_j^0,\,\omega_j=-\frac{k_j^4+3p_j^2}{k_j},\,\exp A_{ij}=\frac{k_i^2k_j^2(k_i-k_j)^2-(k_jp_i-k_ip_j)^2}{k_i^2k_j^2(k_i+k_j)^2-(k_jp_i+k_ip_j)^2}\triangleq a_{ij}\geqslant 0.
	\end{split}
\end{flalign}

Based on the explicit 3-soliton solution, this work aims to provide a systematic and fully analytical study of stem structures generated by resonant interactions in the KPII equation. 
The main objectives and innovations are as follows:
\begin{itemize}
\item To characterize, in a unified framework, the formation, disappearance, and reconnection mechanisms of stem structures in both 2-resonant and 3-resonant interactions, and to reveal their analytical mathematical structures.
\item To derive explicit asymptotic formulas for the trajectories, amplitudes, and velocities of stem structures from the four soliton arms, using and extending the asymptotic analysis method developed in Refs. \cite{jpsj1983,yuan2024,annv2025}.
\item To obtain a complete analytical description of the spatial localization of stem structures, including the positions of their endpoints, their time-dependent lengths, and the locations and evolution of their extrema, under strong, weak, and mixed resonance conditions, consistent with the classification of Ref. \cite{jpsj1983}.
\end{itemize}

The structure of this paper is organized as follows:  
Section 2 presents various types of 2-resonant soliton solutions of the KPII equation and provides a systematic analysis of the properties of the associated stem structures. 
Section 3 extends the discussion to 3-resonant soliton solutions of the KPII equation, offering a comprehensive examination of the corresponding stem characteristics. 
Finally, Section 4 concludes the paper by summarizing the main results, discussing their implications, and proposing potential applications as well as directions for future research.

\section{The stem structure in 2-resonant 3-soliton}\label{sec2}
The phase shift of the 3-soliton solution is denoted as $\Delta_{ij}=\ln a_{ij}$ for $(i,\,j=1,\,2,\,3 \text{ and } i<j)$, while $a_{ij}$ is referred to as the phase shift parameter. 
Different conditions on the phase shift give rise to distinct types of collisions between the three solitons: 
elastic collisions correspond to $|\Delta_{ij}|<+\infty$, while resonance collisions correspond to $|\Delta_{ij}|\to +\infty$ 
(i.e. $a_{ij}\to 0$ or $+\infty$, which we called them as resonance condition). 
In addition, we refer to these two scenarios as strong and weak resonance: 
the strong resonance corresponds to $\Delta_{ij}\to +\infty$ (i.e. $a_{ij}\to +\infty$), and the weak resonance corresponds to $\Delta_{ij}\to -\infty$ (i.e. $a_{ij}\to 0$).

In this section, we narrow our focus to the variable-length stem structure in 2-resonant 3-solitons, occurring when two of $\Delta_{ij}\to \pm\infty$. 
Without loss of generality, we make $\Delta_{13}$ and $\Delta_{23}\to \pm\infty$. 
That is, the phase shift parameters $a_{13}$ and $a_{23}\to +\infty$ or $0$ and $0<a_{12}<+\infty$. 
By combining different resonance conditions, we can obtain various 2-resonant 3-soliton solutions, which will be studied in following three subsections. 

\subsection{Strong 2-resonant case: $a_{13}$ and $a_{23}\to +\infty$}
To obtain the strong 2-resonant 3-soliton, we consider the case where $a_{13}$ and $a_{23}\to +\infty$ in \eqref{3f}. 
It is important to note that we cannot obtain the strong 2-resonant 3-soliton by the limits $a_{13}\to+\infty$ and $a_{23}\to+\infty$ directly, but must obtain it by transformations similar to that in Ref.\ \cite{jpsj1983} (see subsection 4.3 in Ref.\ \cite{jpsj1983}). 
It should be noted that Ref.\ \cite{jpsj1983} reports only three types of transformations, which constitutes an incomplete classification. 
Furthermore, the stem structures associated with the corresponding soliton solutions have not been studied. 
In this subsection, we present the complete set of four transformations and provide a detailed analysis of the stem structure in the first case as a representative example.

By different transformations, the strong 2-resonant 3-solitons have the following cases:

{\bf Case 2.1:} By substituting $\xi_3\to\xi_3-\ln a_{13}-\ln a_{23}$ into Eq.\ \eqref{3f}, and taking the limits $a_{13},\,a_{23}\to +\infty$, we obtain
\begin{flalign}\label{3f2r1}
u=2(\ln f)_{xx},\,	f=1+\exp\xi_1+\exp\xi_2+a_{12}\exp(\xi_1+\xi_2)+a_{12}\exp(\xi_1+\xi_2+\xi_3).
\end{flalign}

{\bf Case 2.2:} Substituting $\xi_1\to\xi_1-\ln a_{13}$ and $\xi_3\to\xi_3-\ln a_{23}$ into Eq.\ \eqref{3f}, and subsequently taking the limits $a_{13},\,a_{23}\to +\infty$, leads to the expression
\begin{flalign}\label{3f2r2}
u=2(\ln f)_{xx},\,	f=1+\exp\xi_2+\exp(\xi_2+\xi_3)+a_{12}\exp(\xi_1+\xi_2+\xi_3).
\end{flalign}

{\bf Case 2.3:} When $\xi_2\to\xi_2-\ln a_{23}$ and $\xi_3\to\xi_3-\ln a_{13}$ are substituted into Eq.\ \eqref{3f}, and the limits $a_{13},\,a_{23}\to +\infty$ are taken, we can derive
\begin{flalign}\label{3f2r3}
u=2(\ln f)_{xx},\,	f=1+\exp\xi_1+\exp(\xi_1+\xi_3)+a_{12}\exp(\xi_1+\xi_2+\xi_3).
\end{flalign}

{\bf Case 2.4:} Under the condition $\xi_1\to\xi_1-\ln a_{13}$ and $\xi_2\to\xi_2-\ln a_{23}$ into Eq.\ \eqref{3f}, and in the limiting case where $a_{13},\,a_{23}\to +\infty$, Eq.~\eqref{3f} simplifies to yield
\begin{flalign}\label{3f2r4}
u=2(\ln f)_{xx},\,	f=1+\exp\xi_3+\exp(\xi_1+\xi_3)+\exp(\xi_2+\xi_3)+a_{12}\exp(\xi_1+\xi_2+\xi_3).
\end{flalign}

\begin{rk}
When \( a_{12} \to +\infty \) and either \( a_{13} \to +\infty \) or \( a_{23} \to +\infty \), one can derive alternative expressions through transformations analogous to those employed in the preceding cases. 
However, these resulting formulae do not introduce any essential new features; 
rather, they are equivalent to those in Cases~2.1--2.4, differing only by a permutation of the subscripts in \( \xi_j \) and $a_{ij}$. 
In fact, Cases~2.1--2.3 correspond precisely to Eqs.~(4.16)--(4.23) in Ref.~\cite{jpsj1983}, after interchanging \( \xi_2 \) and \( \xi_3 \).
\end{rk}

\begin{rk}
Cases~2.1--2.4 are generally applicable to soliton equations whose tau functions take the form of Eq.~\eqref{3f}, regardless of the specific form of \( \xi_j \).
\end{rk}

Since there is no essential difference among the four cases under consideration, similar results can be derived using the same analytical approach. 
Therefore, for the sake of brevity, we focus exclusively on Case 2.1 for a detailed analysis.

To ensure that \( a_{13}\) and \( a_{23} \to +\infty \), the parameters must satisfy either
\[p_1 = \frac{k_1 (k_1 k_3 + k_3^2 + p_3)}{k_3}, \quad p_2 = -\frac{k_2 (k_2 k_3 + k_3^2 - p_3)}{k_3},\]
or
\[p_1 = -\frac{k_1 (k_1 k_3 + k_3^2 - p_3)}{k_3}, \quad p_2 = \frac{k_2 (k_2 k_3 + k_3^2 + p_3)}{k_3}.\]
Without loss of generality, we consider only the former case in this subsection. Under this condition, the expression for \( a_{12} \) becomes
$a_{12} = \frac{(k_1 + k_3)(k_2 + k_3)}{k_3(k_1 + k_2 + k_3)}$, which must satisfy \( 0 < a_{12} < +\infty \). 
Following the same step-by-step analytical procedure as in Refs.~\cite{yuan2024,annv2025}, we arrive at the following proposition.
\begin{prop}\label{prop2.1}
The asymptotic forms of the strong 2-resonant 3-soliton \eqref{3f2r1} with 
$p_1 = \frac{k_1 \left( k_1 k_3 + k_3^2 + p_3 \right)}{k_3}, \, p_2 = -\frac{k_2 \left( k_2 k_3 + k_3^2 -p_3 \right)}{k_3}$ are as following:

\noindent Before collision ($t\to-\infty$):
\begin{flalign}\label{3s2rasy03}
\begin{split}
y\to -\infty,\quad &S_{1+3}:\,\,u\sim \widehat{u_{1+3}},\quad S_2:\quad u\sim u_2,\\
y\to +\infty,\quad &S_1:\quad u\sim u_1,\quad S_{2+3}:\,\,u\sim \widehat{u_{2+3}}.
\end{split}
\end{flalign}

\noindent After collision ($t\to+\infty$):
\begin{flalign}\label{3s2rasy04}
\begin{split}
y\to -\infty,\quad &S_{1+3}:\,\,u\sim \widehat{u_{1+3}},\quad S_2:\quad u\sim \widehat{u_2},\\
y\to +\infty,\quad &S_1:\quad u\sim \widehat{u_1}, \quad S_{2+3}:\,\,u\sim \widehat{u_{2+3}}.
\end{split}
\end{flalign}

\noindent The stem structures:
\begin{equation}\label{3s2rstem02}
t\to-\infty,\quad S_{1+2+3}:\, u\sim \widehat{u_{1+2+3}};\qquad t\to+\infty,\quad S_3:\,u\sim u_3.
\end{equation}

Here, $S_j,\,S_{i+j},\,S_{i+j+k}$ are the soliton arms and stem structures of the soliton solutions given by Table \ref{tab:t1}, 
and $u_j,\,u_{i+j},\,u_{i+j+k}$, $\widehat{u_j},\,\widehat{u_{i+j}},\,\widehat{u_{i+j+k}}$ are given by
\begin{flalign}\label{uj01}
\begin{split}
&u_j= \frac{k_j^2}{2}\sech^2\frac{\xi_j}{2},\,\widehat{u_j}= \frac{k_j^2}{2}\sech^2\frac{\xi_j+\ln a_{12}}{2},\,u_{i+ j}= \frac{(k_i+ k_j)^2}{2}\sech^2\frac{\xi_i+ \xi_j}{2},\\
&\widehat{u_{i+ j}}=\frac{(k_i+ k_j)^2}{2}\sech^2\frac{\xi_i+ \xi_j+\ln a_{12}}{2},\,\widehat{u_{1+2+3}}= \frac{(k_1+k_2+k_3)^2}{2}\sech^2\frac{\xi_1+\xi_2+\xi_3+\ln a_{12}}{2}.
\end{split}
\end{flalign}
\end{prop}

By comparing Proposition~\ref{prop2.1} with Eq.~(4.23) in Ref.~\cite{jpsj1983}, we find that Eq.~\eqref{3s2rasy03} corresponds to Eq.~(4.23) under the coordinate transformation \( y \to -y \). However, Eq.~\eqref{3s2rasy04} is not provided in Ref.~\cite{jpsj1983}.
This discrepancy arises because the asymptotic form Eq.\ (4.23) presented in Ref.~\cite{jpsj1983} does not take into account the effect of phase shift before and after the interaction. 
In reality, due to the influence of \( a_{12} \), the soliton arms \( S_1 \) and \( S_2 \) experience a phase shift before and after their interaction. 
This difference is reflected in the expressions \( u_j \) and \(\widehat{u_j}\), where \( u_j \) denotes the case without the presence of \( a_{12} \), and \(\widehat{u_j} \) corresponds to the case where \( a_{12} \) is included.
In addition, because $a_{13}$ and $a_{23}$ meet the strong resonance conditions, strong resonances occur between $ S_1$ and $S_3$, $S_2$ and $S_3$ to generate $S_{1+3}$ and $S_{2+3}$, and $S_{1+2+3}$ is obtained by further resonance.

\begin{figure}[ht]
\centering
\subfigure[$t=-2$]{\includegraphics[height=4cm,width=5cm]{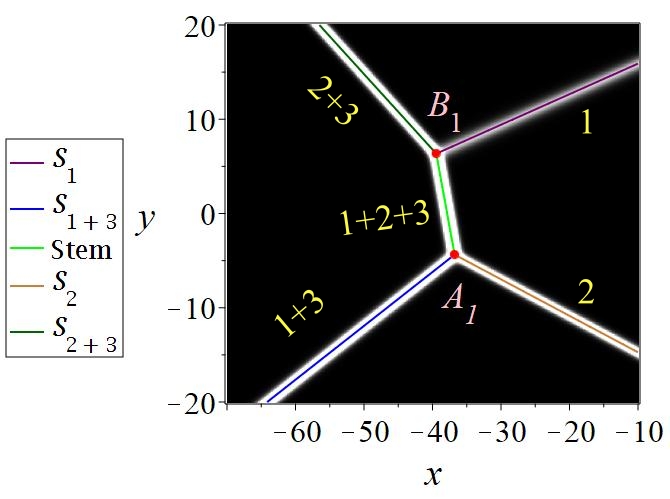}}
\subfigure[$t=0$]{\includegraphics[height=4cm,width=4cm]{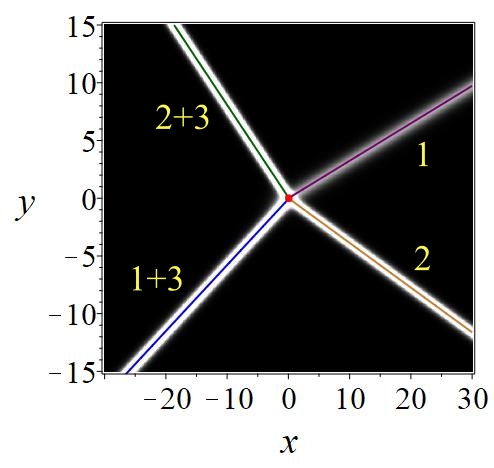}}
\subfigure[$t=1$]{\includegraphics[height=4cm,width=4cm]{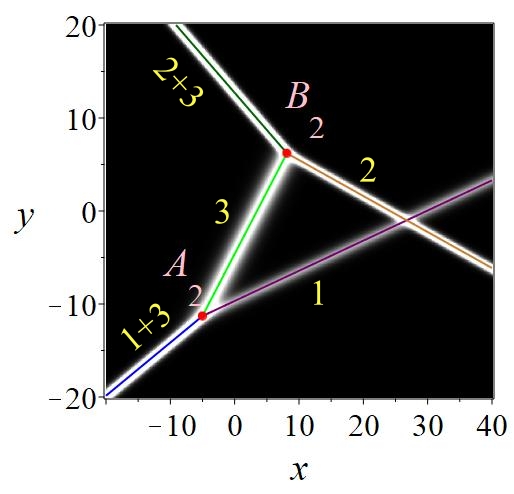}}
\vspace{-0.6\baselineskip}
\caption{The density plots of the strong 2-resonant 3-soliton \eqref{3f2r1} with 
$k_1=-1,\,k_2=-2,\,k_3=-\frac{4}{3},\,p_1 = \frac{k_1 \left( k_1 k_3 + k_3^2 + p_3 \right)}{k_3}, \, p_2 = -\frac{k_2 \left( k_2 k_3 + k_3^2 -p_3 \right)}{k_3},\,p_3=1$. 
The lines are the trajectories of the arms and stem structures, and the red points are the endpoints of the variable-length stem structures.
}\label{fig2-1}
\end{figure}

The dynamics involving the evolution of four soliton arms and the associated loss and fission of the stem structure are depicted in Fig.~\ref{fig2-1}. 
As $t \to -\infty$, the stem structure $S_{1+2+3}$ links two distinct pairs of V-shaped solitons: ($S_1$, $S_{2+3}$) and ($S_2$, $S_{1+2}$). 
With the progression of time, the stem structure gradually shortens and eventually vanishes near $t = 0$. 
At this critical moment, all four soliton arms ($S_1$, $S_2$, $S_{1+3}$, $S_{2+3}$) converge, resulting in a transformation of the initial soliton pairs into new V-shaped configurations: ($S_1$, $S_{1+3}$) and ($S_2$, $S_{2+3}$). 
As time continues to advance ($t \to +\infty$), a re-emergent stem structure $S_{1+2+3}$ forms and progressively elongates, once again linking the two reconfigured soliton pairs. 
This process, whereby the soliton arms reorganize to form new V-shaped structures, is referred to as soliton reconnection. 
The amplitudes and velocities of the four arms and the two stem structures are summarized in Table~\ref{tab:t1}, where their trajectories are described by
\begin{flalign}\label{l01}
\begin{split}
&\boldsymbol{l_j:}\,\xi_j=0,\quad \boldsymbol{\widehat{l_j}:}\,\xi_j+\ln a_{12}=0,\quad \boldsymbol{l_{i+ j}:}\, \xi_i+ \xi_j=0,\\
&\boldsymbol{\widehat{l_{i+ j}}:}\, \xi_i+ \xi_j+\ln a_{12}=0,\,\boldsymbol{\widehat{l_{1+2+3}}:}\,\xi_1+\xi_2+\xi_3+\ln a_{12}=0.
\end{split}
\end{flalign}

It can be readily verified that, as $t \to -\infty$, the trajectories of $S_1$, $S_{2+3}$, and $S_{1+2+3}$ intersect at a common point, and likewise, the trajectories of $S_2$, $S_{1+2}$, and $S_{1+2+3}$ also intersect at a single point. 
Similarly, as $t \to +\infty$, the trajectories of $S_1$, $S_3$, and $S_{1+3}$ converge at one point, while those of $S_2$, $S_3$, and $S_{2+3}$ intersect at another. 
These intersection properties serve as further validation for the spatial localization of the stem structures described by Eq.~\eqref{3s2rstem02}. 
Fig.~\ref{fig2-1} presents the soliton trajectories of $u$ at different time slices, where the background shading corresponds to the density distribution. 
By analytically determining the intersection points of these characteristic trajectories, the endpoints of the variable-length stem structures can be explicitly identified as follows:
\begin{flalign}\label{endpoints1}
\begin{split}
A_1\,&\Bigg(\bigg(k_3^2+4k_1k_2+4k_2k_3-2p_3+\frac{4k_2p_3-4p_3k_1}{k_3}-\frac{3p_3^2}{k_3^2}\bigg)t-\frac{(k_1k_3+k_3^2+p_3)\ln a_{12}}{k_3(k_1+k_2+k_3)(k_2+k_3)},\,\\
&\frac{\ln a_{12}}{(k_1+k_2+k_3)(k_2+k_3)}+\bigg(\frac{6p_3}{k_3}+4k_1-4k_2+2k_3\bigg)t\Bigg),\\
B_1\,&\Bigg(\bigg(k_3^2+4k_1k_2+4k_1k_3+2p_3-\frac{4k_1p_3-4p_3k_2}{k_3}-\frac{3p_3^2}{k_3^2}\bigg)t-\frac{(k_2k_3+k_3^2-p_3)\ln a_{12}}{k_3(k_1+k_2+k_3)(k_1+k_3)},\,\\
&-\frac{\ln a_{12}}{(k_1+k_2+k_3)(k_1+k_3)}+\bigg(\frac{6p_3}{k_3}-4k_1+4k_2+2k_3\bigg)t\Bigg),\\
A_2\,&\Bigg(\frac{p_3\ln a_{12}}{k_1k_3(k_1+k_3)}-\bigg(\frac{4k_1k_3p_3+3p_3^2}{k_3^2}-k_3^2+2p_3\bigg)t,\,-\frac{\ln a_{12}}{k_1(k_1+k_3)} +\bigg(\frac{6p_3}{k_3}+4k_1+2k_3\bigg)t \Bigg),\,\\
B_2\,&\Bigg(\frac{-p_3\ln a_{12}}{k_2k_3(k_2+k_3)}+\bigg(\frac{4k_2k_3p_3-3p_3^2}{k_3^2}+k_3^2+2p_3\bigg)t,\,\frac{\ln a_{12}}{k_2(k_2+k_3)} +\bigg(\frac{6p_3}{k_3}-4k_2-2k_3\bigg)t \Bigg).
\end{split}
\end{flalign}
In this setting, point $A_1$ represents the intersection between the trajectories $\widehat{l_{2+3}}$ and $l_1$, while point $B_1$ corresponds to the intersection of $\widehat{l_{1+2}}$ and $l_2$. 
Similarly, point $A_2$ identifies the intersection of $l_3$ with $\widehat{l_{1+3}}$, and point $B_2$ denotes the intersection between $l_3$ and $\widehat{l_{2+3}}$. 
These intersection points are also annotated in Fig.~\ref{fig2-1}. 
Based on these geometric configurations, the lengths of the stem structures can be determined as follows:
\begin{flalign}\label{3length-1}
\begin{split}
&|A_1B_1|=\left|4t+\frac{\ln a_{12}}{k_3(k_1+k_3)(k_2+k_3)}\right|\sqrt{k_3^2+(k_3(k_1-k_2)+p_3)^2},\,\text{where}\,t\ll 0,\\
&|A_2B_2|=\sqrt{k_3^2+p_3^2}\left|\frac{\ln a_{12}}{k_1k_3(k_1+k_3)}+\frac{\ln a_{12}}{k_2k_3(k_2+k_3)}-\frac{4(k_1+k_2+k_3)t}{k_3} \right|,\,\text{where}\,t\gg 0.
\end{split}
\end{flalign}

From the above equation, it is evident that both $|A_1B_1|$ and $|A_2B_2|$ exhibit a linear dependence on $t$ in the regime where $|t| \gg 0$.  
However, it is important to emphasize that the evolution of the soliton and its associated stem structure near $t = 0$ is highly intricate, rendering the derivation of an explicit length formula in this region infeasible.  
As a result, Eq.~\eqref{3length-1} is not valid in the vicinity of $t = 0$.  
Similarly, we are unable to precisely determine how the endpoints of the stem structure transition from $A_1$ and $B_1$ to $A_2$ and $B_2$ in the vicinity of $t = 0$.  
As can also be seen from Eqs.\ \eqref{l01} and \eqref{endpoints1}, there exists no specific time $t$ at which all four soliton arms intersect at a single point.  
Therefore, although the two endpoints depicted in Fig.~\ref{fig2-1} (b) appear to coincide due to their close proximity, they are in fact distinct and do not overlap.
The same reasoning extends to the remaining resonant solutions analyzed in this section. 

By comparing with Fig.~6 in Ref.~\cite{jpsj1983}, it can be observed that the stem structure region is not depicted in the figure, and the asymptotic form does not reflect the behavior as $t \to \pm\infty$.  
In fact, the asymptotic representation and the corresponding graph provided in Section~4.3 of Ref.~\cite{jpsj1983} are more closely associated with the behavior near $t = 0$.  
In contrast, the present work derives asymptotic forms for $t \to \pm\infty$ (as presented in Proposition~\ref{prop2.1}) and offers a detailed analysis of the stem structure, thereby extending and refining the results of Ref.~\cite{jpsj1983}.  

\begin{figure}[h!tb]
\centering
\subfigure[$t=-2$]{\includegraphics[height=3cm,width=3cm]{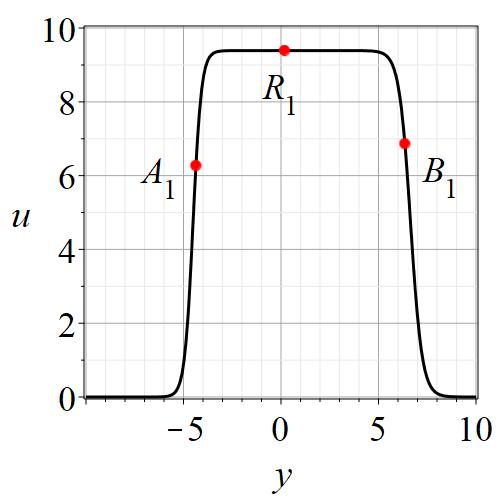}}
\subfigure[$t=1$]{\includegraphics[height=3cm,width=3cm]{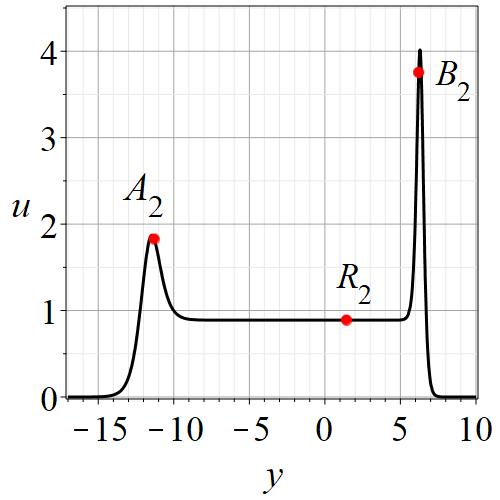}}
\subfigure[]{\includegraphics[height=3cm,width=3cm]{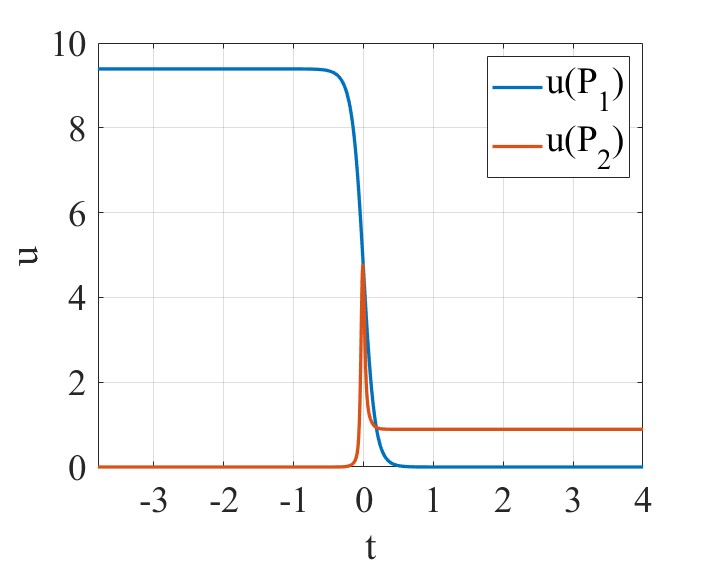}}
\subfigure[$t=-2$]{\includegraphics[height=3cm,width=3cm]{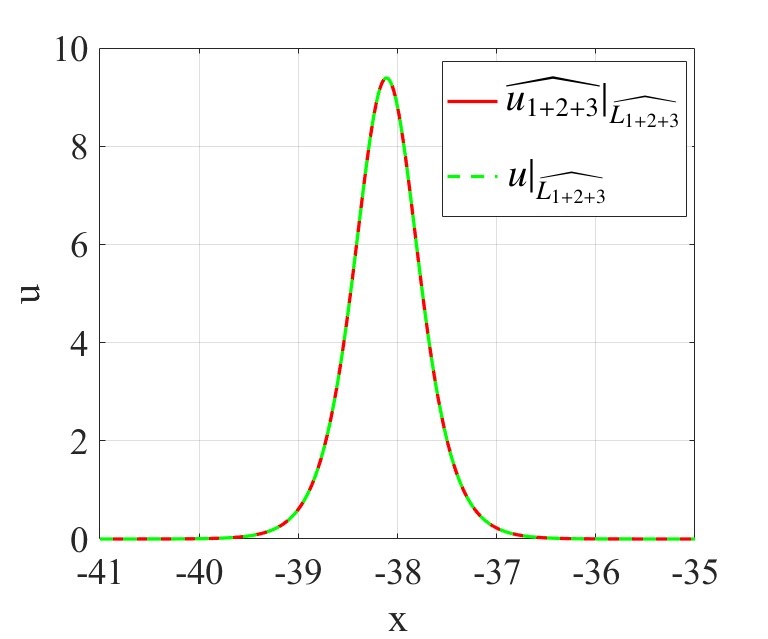}}
\subfigure[$t=1$]{\includegraphics[height=3cm,width=3cm]{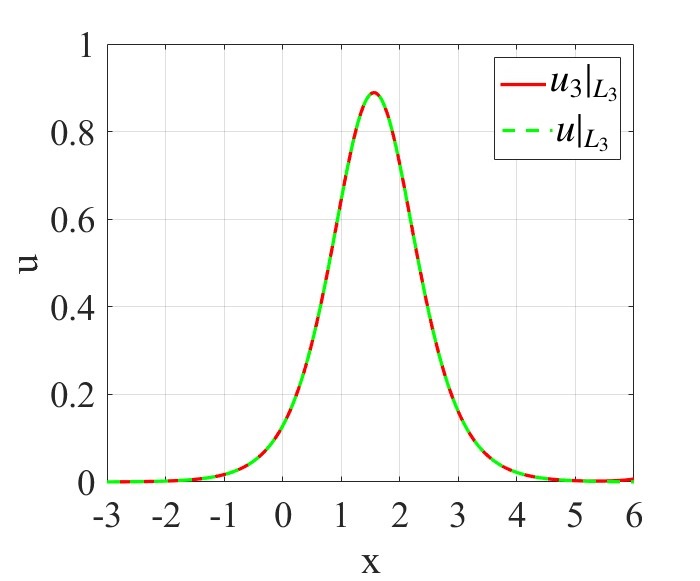}}
\vspace{-0.6\baselineskip}
\caption{Parameters: $k_1=-1,\,k_2=-2,\,k_3=-\frac{4}{3},\,p_1 = \frac{k_1 \left( k_1 k_3 + k_3^2 + p_3 \right)}{k_3}, \, p_2 = -\frac{k_2 \left( k_2 k_3 + k_3^2 -p_3 \right)}{k_3},\,p_3=1$.
(a) The cross-sectional curve $u|_{\widehat{l_{1+2+3}}}$ given by \eqref{cross3s01}; 
(b) The cross-sectional curve $u|_{l_3}$ given by \eqref{cross3s01}; 
(c) The amplitude evolution curves \eqref{up01}; 
(d) The cross-sectional curves $u|_{L_{1+2+3}}$ and $\widehat{u}_{1+2+3}|_{L_{1+2+3}}$; 
(e) The cross-sectional curves $u|_{L_3}$ and $u_3|_{L_3}$.
}\label{fig2-2}
\end{figure}

We now proceed to analyze the amplitudes of variable-length stem structures.  
The cross-sectional profiles of the 3-soliton solution~\eqref{3f}, determined using Eqs.~\eqref{3f2r2} along $\widehat{l_{1+2+3}}$ and $l_3$ as illustrated in Fig.~\ref{fig2-2} (a) and (b), are given by
\begin{flalign}\label{cross3s01}
\begin{split}
&u|_{\widehat{l_{1+2+3}}}=\frac{2a_{12}k_1^2\e^{\theta_1+2\theta_2}+2a_{12}k_2^2\e^{2\theta_1+\theta_2}+(2a_{12}(k_1+k_2)^2-2(k_1-k_2)^2)\e^{\theta_1+\theta_2}+2k_1^2\e^{\theta_1}
+2k_2^2\e^{\theta_2}}{(1+\e^{\theta_1}+\e^{\theta_2}+a_{12}\e^{\theta_1+\theta_2})^2},\\
&u|_{l_3}=\frac{4a_{12}k_1^2\e^{\theta_3+2\theta_4}+4a_{12}k_2^2\e^{2\theta_3+\theta_4}+(4a_{12}(k_1+k_2)^2+2(k_1-k_2)^2)\e^{\theta_3+\theta_4}+2k_1^2\e^{\theta_3}
+2k_2^2\e^{\theta_4}}{(1+\e^{\theta_3}+\e^{\theta_4}+2a_{12}\e^{\theta_3+\theta_4})^2},
\end{split}
\end{flalign}
where, 
\begin{flalign*}
&\theta_1=\frac{p_1(k_2+k_3)-k_1(p_2+p_3)}{k_1+k_2+k_3}y+\Bigg(\frac{k_1(k_2^3 - k_1^2k_2 - k_1^2k_3 + k_3^3)}{k_1 + k_2 + k_3}
+\frac{3(k_1^2p_3^2 - k_3^2p_1^2) }{k_1k_3(k_1 + k_2 + k_3)}\\
&+\frac{3(k_1^2p_2^2 - k_2^2p_1^2) }{k_1k_2(k_1 + k_2 + k_3)}\Bigg)t-\frac{k_1\ln a_{12}}{k_1 + k_2 + k_3},\\
&\theta_2=\frac{p_2(k_1+k_3)-k_2(p_1+p_3)}{k_1+k_2+k_3}y+\Bigg(\frac{k_2(k_1^3 - k_1k_2^2 - k_2^2k_3 + k_3^3)}{k_1 + k_2 + k_3}
+\frac{3(k_2^2p_3^2 - k_3^2p_2^2) }{k_2k_3(k_1 + k_2 + k_3)}\\
&+\frac{3(k_2^2p_1^2 - k_1^2p_2^2) }{k_1k_2(k_1 + k_2 + k_3)}\Bigg)t-\frac{k_2\ln a_{12}}{k_1 + k_2 + k_3},\\
&\theta_3=\frac{(k_3p_1-k_1p_3)y}{k_3} + \Bigg(k_1k_3^2 -k_1^3 + \frac{3k_1p_3^2}{k_3^2} - \frac{3p_1^2}{k_1} \Bigg)t,\\
&\theta_4=\frac{(k_3p_2-k_2p_3)y}{k_3} + \Bigg(k_2k_3^2 -k_2^3 + \frac{3k_2p_3^2}{k_3^2} - \frac{3p_2^2}{k_2} \Bigg)t.
\end{flalign*}

Due to the complexity of the calculations, the extrema of the cross-sectional curve~\eqref{cross3s01} cannot be obtained analytically. 
Instead, for given parameters ($k_j$, $p_j$), and $t$, numerical values of the extrema and their corresponding amplitudes along the stem structure can be determined. 
For instance, when the parameters are the same as those in Fig.~\ref{fig2-1} (a), the extremum occurs at the point $(0.187$,\, $9.389)$, with the extreme value $9.389 \approx \frac{(k_1 + k_2 + k_3)^2}{2} = \frac{169}{18}$, as indicated by the red point $R_1$ in Fig.~\ref{fig2-2} (a). 
Similarly, when the parameters correspond to Fig.~\ref{fig2-1} (c), the extremum is located at $(1.440$,\, $0.889)$, with the extreme value $0.889 \approx \frac{k_3^2}{2} = \frac{8}{9}$, as shown by the red point $R_2$ in Fig.~\ref{fig2-2} (b). 
However, these extrema only yield instantaneous numerical values and do not capture the temporal evolution of the amplitude, which poses difficulties for further analytical exploration. 
To address this limitation, we consider the time-dependent amplitude at the midpoint of the stem structures, given by the following expressions:
\begin{flalign} \label{up01}
\begin{split}
&u(P_1)=\frac{F_1+279897800}{6084\left(\alpha_5\e^{\frac{112t}{9}}+\alpha_4\e^{\frac{80t}{9}}+\frac{35\,\alpha_3}{26}\e^{\frac{64t}{3}}+70\right)^2},\\
&u(P_2)=\frac{3719825200\,\e^{\frac{182t}{9}}F_2}{1863225\left(\alpha_9\e^{\frac{520t}{9}}+\frac{26\,\alpha_{11}}{35}\e^{\frac{182t}{9}}+\frac{35\,\alpha_6}{13}\e^{78t}+26\right)^2},
\end{split}
\end{flalign}
where 
$F_1=2293200\,\alpha_1\e^{\frac{272t}{9}}+425880\,\alpha_2\e^{\frac{304t}{9}}+6604780\,\alpha_3\e^{\frac{64t}{3}}
+5157880\,\alpha_4\e^{\frac{80t}{9}}+4022200\,\alpha_5\e^{\frac{112t}{9}}$, 
$F_2=\alpha_6\e^{\frac{520t}{9}}+\frac{35525\,\alpha_7}{116792}\e^{\frac{1040t}{9}}+\frac{595\,\alpha_8}{1123}\e^{78t}
+\frac{117\,\alpha_9}{1123}\e^{\frac{338t}{9}}+\frac{42875\,\alpha_{10}}{379574}\e^{\frac{1222t}{9}}+\frac{1521\,\alpha_{11}}{78610}$, 
$\alpha_1=35^{\frac{43}{910}}26^{\frac{867}{910}}$,\,$\alpha_2=35^{\frac{851}{910}}26^{\frac{59}{910}}$, 
$\alpha_3=35^{\frac{149}{455}}26^{\frac{306}{455}}$,\,$\alpha_4=35^{\frac{131}{182}}26^{\frac{51}{182}}$, 
$\alpha_5=35^{\frac{79}{130}}26^{\frac{51}{130}}$,\,$\alpha_6=35^{\frac{169}{280}}26^{\frac{111}{280}}$, 
$\alpha_7=35^{\frac{149}{280}}26^{\frac{131}{280}}$, $\alpha_8=35^{\frac{39}{140}}26^{\frac{101}{140}}$, 
$\alpha_9=35^{\frac{13}{14}}26^{\frac{1}{14}}$,\,$\alpha_{10}=35^{\frac{29}{140}}26^{\frac{111}{140}}$,\,
$\alpha_{11}=35^{\frac{27}{40}}26^{\frac{13}{40}}$,\,
and $P_1,\,P_2$ are the midpoints of $A_1B_1$ and $A_2B_2$, respectively. 

The amplitude evolution curves of $u(P_1)$ and $u(P_2)$ are shown in Fig.~\ref{fig2-2} (c). 
As observed in the figure, the amplitudes given by~\eqref{up01} exhibit rapid and complex variations near $t = 0$, and gradually stabilize as $|t| \gg 0$. 
This behavior arises because, around $t = 0$, the four arms are in close proximity and interact strongly, making the stem structure indistinct. 
As a result, studying the properties of the stem in this regime offers limited insight. 
Notably, we have
\vspace{-0.5\baselineskip}
\[\lim_{t \to -\infty} u(P_1) = \frac{169}{18} = \frac{(k_1 + k_2 + k_3)^2}{2}, \quad 
\lim_{t \to +\infty} u(P_2) = \frac{8}{9} = \frac{k_3^2}{2},\vspace{-0.5\baselineskip}\]
which justifies using $u(P_1)$ and $u(P_2)$ as reliable approximations for the amplitudes of the stem structures $S_{1+2+3}$ and $S_3$, respectively. 
Furthermore, we consider the vertical plane passing through point $P_1$ and perpendicular to the direction $\widehat{l_{1+2+3}}$, defined by
\vspace{-0.5\baselineskip}
\[L_{1+2+3}:\quad (p_1 + p_2 + p_3)x - (k_1 + k_2 + k_3)y - \left[(p_1 + p_2 + p_3)x_{P_1} - (k_1 + k_2 + k_3)y_{P_1}\right] = 0,\vspace{-0.5\baselineskip}\]
where $x_{P_1}$ and $y_{P_1}$ denote the $x$- and $y$-coordinates of the point $P_1$, respectively. 
Fig.~\ref{fig2-2} (d) shows the intersection curves of $u$ and $\widehat{u_{1+2+3}}$ with the plane $L_{1+2+3}$, denoted by $u|_{L_{1+2+3}}$ and $\widehat{u_{1+2+3}}|_{L_{1+2+3}}$, respectively. 
Similarly, the vertical plane passing through point $P_2$ and perpendicular to the direction of $L_3$ is given by
\vspace{-0.5\baselineskip}
\[L_3:\quad p_3x - k_3y - (p_3x_{P_2} - k_3y_{P_2}) = 0,\]
where $x_{P_2}$ and $y_{P_2}$ are the $x$- and $y$-coordinates of point $P_2$, respectively. 
Fig.~\ref{fig2-2}~(e) displays the intersection curves of $u$ and $u_3$ with $L_3$, denoted by $u|_{L_3}$ and $u_3|_{L_3}$, respectively. 

As shown in Fig.~\ref{fig2-2} (d) and (e), these curves nearly coincide. 
This observation further supports the validity of using $u_3$ and $\widehat{u_{1+2+3}}$ as approximations of $u$ in the limits $t \to -\infty$ and $t \to +\infty$, respectively.

In addition, it should be noted that for the resonant 3-soliton solution, the asymptotic behavior is influenced by the coefficients of $x$, $y$, and $t$ in $\xi_j$. 
This implies that different parameter choices may lead to distinct asymptotic structures. 
For instance, by selecting the parameters 
$k_1 = -1$, $k_2 = \frac{3}{2}$, $k_3 = 2$, and $p_3 = 1$, we obtain an asymptotic configuration that differs from that described in Proposition~\ref{prop2.1}, as shown below:

Before collision ($t\to-\infty$):
\begin{flalign}
\begin{split}
x\to -\infty,\quad &S_1:\quad  u\sim \widehat{u_1},\quad S_2:\quad u\sim u_2;\\
x\to +\infty,\quad &S_3:\quad u\sim u_3,\quad S_{1+2+3}:\,\,u\sim \widehat{u_{1+2+3}}.
\end{split}
\end{flalign}

After collision ($t\to+\infty$):
\begin{flalign}
\begin{split}
x\to -\infty,\quad &S_1:\quad u\sim u_1,\quad S_2:\quad u\sim \widehat{u_2};\\
x\to +\infty,\quad &S_3:\quad u\sim u_3,\quad S_{1+2+3}:\,\,u\sim \widehat{u_{1+2+3}}.
\end{split}
\end{flalign}

The stem structures:
\begin{flalign}
\begin{split}
t\to-\infty,\quad S_{1+3}:\quad u\sim \widehat{u_{1+3}};\quad t\to+\infty,\quad S_{2+3}:\, u\sim \widehat{u_{2+3}}.
\end{split}
\end{flalign}

\begin{figure}[h!tb]
\centering
\subfigure[$t=-2$]{\includegraphics[height=4cm,width=5.5cm]{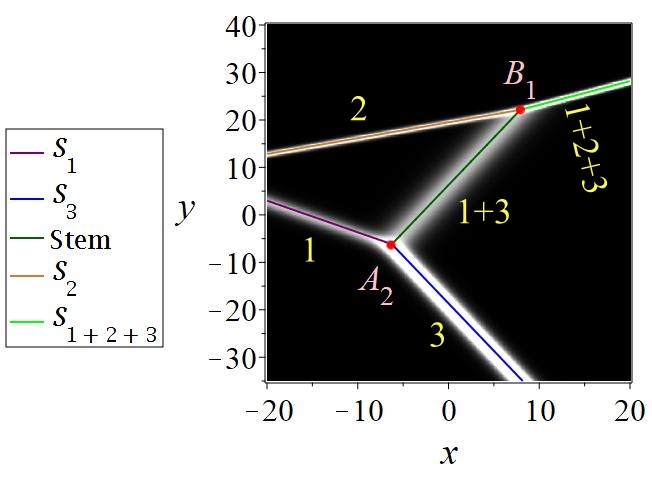}}
\subfigure[$t=4$]{\includegraphics[height=4cm,width=4cm]{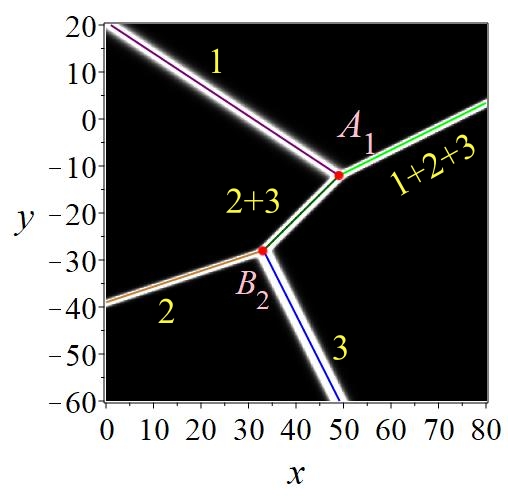}}
\vspace{-0.6\baselineskip}
\caption{The density plots of the strong 2-resonance 3-soliton \eqref{3f2r1} with 
$k_1=-1,\,k_2=\frac{3}{2},\,k_3=2,\,p_1 = \frac{k_1 \left( k_1 k_3 + k_3^2 + p_3 \right)}{k_3}, \, p_2 = -\frac{k_2 \left( k_2 k_3 + k_3^2 -p_3 \right)}{k_3},\,p_3=1$. 
The lines are the trajectories of the arms and stem structures, and the points are the endpoints of the variable length stem structures.
}\label{fig2-3}
\end{figure}

By comparing the above expression with Proposition~\ref{prop2.1}, it is evident that the four soliton arms are not identical, and the corresponding stem structures also differ. 
The density plot and the corresponding soliton trajectories are shown in Fig.~\ref{fig2-3}, where the red points indicate the endpoints of the stem structure. 
Since the analysis of the stem structure in this case follows a similar methodology, further details are omitted here.

Moreover, the asymptotic forms of resonant solitons, the trajectory equations of the soliton arms, and the endpoint formulas for the stem structures in 
Cases~2.2-2.4 (Eqs.~\eqref{3f2r2}-\eqref{3f2r4}) are provided in the Appendix, including the corresponding figures. 

\subsection{Weak 2-resonant case: $a_{13}$ and $a_{23} \to 0$}
In this subsection, we investigate the weak 2-resonant 3-soliton case under the conditions $a_{13},\, a_{23} \to 0$ and $0 < a_{12} < +\infty$. 
As a consequence, we obtain the weak 2-resonant 3-soliton solution is given by the following expression:
\begin{flalign}\label{3f2r5}
u=2(\ln f)_{xx},\,f=1+\exp\xi_1+\exp\xi_2+\exp\xi_3+a_{12}\exp(\xi_2+\xi_3).
\end{flalign}
To satisfy the conditions $a_{13},\, a_{23} \to 0$ and $0 < a_{12} < +\infty$, the parameters must be chosen as either
\[p_1 = -\frac{k_1 (k_1 k_3 - k_3^2 - p_3)}{k_3}, \quad p_2 = \frac{k_2 (k_2 k_3 - k_3^2 + p_3)}{k_3},\]
or
\[p_1 = \frac{k_1 (k_1 k_3 - k_3^2 + p_3)}{k_3}, \quad p_2 = -\frac{k_2 (k_2 k_3 - k_3^2 - p_3)}{k_3}.\]
Since these two cases differ only by a transformation in the $y$-coordinate, which does not affect the essential structure of the solution, we focus solely on the first case for generality. 
Under this choice, we obtain $a_{12} = -\frac{(k_1 - k_3)(k_2 - k_3)}{k_3(k_1 + k_2 - k_3)} > 0$, 
which implies that $k_3(k_1 - k_3)(k_2 - k_3)(k_1 + k_2 - k_3) < 0$.
To investigate the weak 2-resonant 3-soliton solution described by Eq.~\eqref{3f2r5}, we analyze its asymptotic behavior. 
Using the progressive analysis method as previous subsection, we obtain the following proposition:
\begin{prop}\label{prop2.2}
The asymptotic forms of the weak 2-resonant 3-soliton \eqref{3f2r5} with 
$p_1 = -\frac{k_1 \left( k_1 k_3 - k_3^2 - p_3 \right)}{k_3},\, p_2 = \frac{k_2 \left( k_2 k_3 - k_3^2 +p_3 \right)}{k_3}$ are as following:

Before collision ($t\to-\infty$):
\begin{flalign}\label{3s2rasy01}
\begin{split}
y\to -\infty,\quad &S_1:\qquad u\sim u_1,\quad S_2:\quad u\sim\widehat{u_2};\\
y\to +\infty,\quad &S_{1+2-3}:\,\,u\sim \widehat{u_{1+2-3}},\,S_3:\quad u\sim u_3.
\end{split}
\end{flalign}

After collision ($t\to+\infty$):
\begin{flalign}\label{3s2rasy02}
\begin{split}
y\to -\infty,\quad &S_1:\qquad u\sim \widehat{u_1},\quad S_2:\, u\sim u_2;\\
y\to +\infty,\quad &S_{1+2-3}:\,\,u\sim \widehat{u_{1+2-3}},\,S_3:\, u\sim u_3.
\end{split}
\end{flalign}

The stem structures:
\begin{flalign}\label{3s2rstem01}
\begin{split}
t\to-\infty,\quad S_{1-3}:\, u\sim u_{1-3};\quad t\to+\infty,\quad S_{2-3}:\, u\sim u_{2-3}.
\end{split}
\end{flalign}

Here, the relevant formulas are given by Eq.\ \eqref{uj01} and
\begin{flalign}\label{uj02}
\begin{split}
&u_{i- j}= \frac{(k_i-k_j)^2}{2}\sech^2\frac{\xi_i-\xi_j}{2},\,\widehat{u_{1+2-3}}= \frac{(k_1+k_2-k_3)^2}{2}\sech^2\frac{\xi_1+\xi_2-\xi_3+\ln a_{12}}{2}.
\end{split}
\end{flalign}
\end{prop}

\begin{figure}[h!tb]
\centering
\subfigure[$t=-2$]{\includegraphics[height=4cm,width=5.3cm]{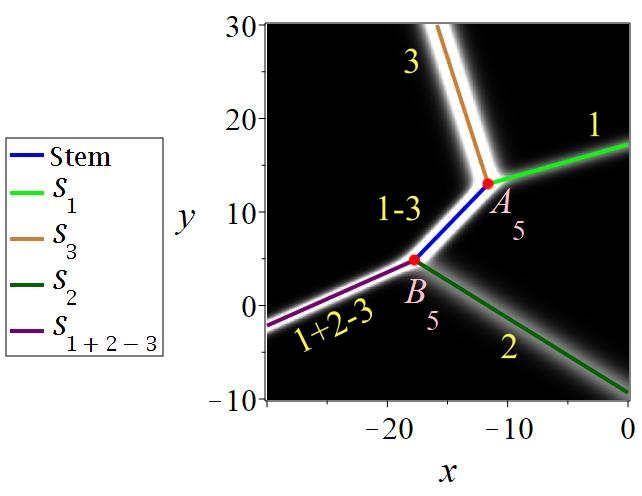}}
\subfigure[$t=0$]{\includegraphics[height=4cm,width=4cm]{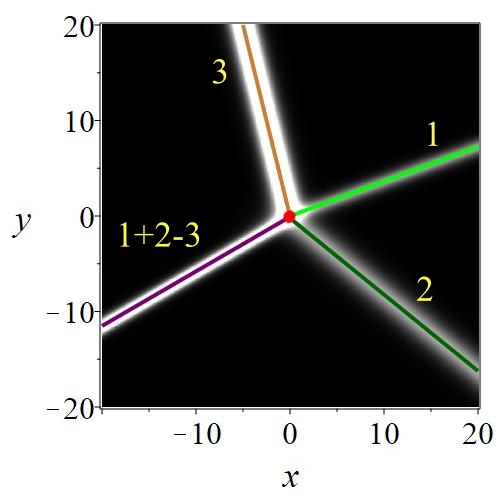}}
\subfigure[$t=2$]{\includegraphics[height=4cm,width=4cm]{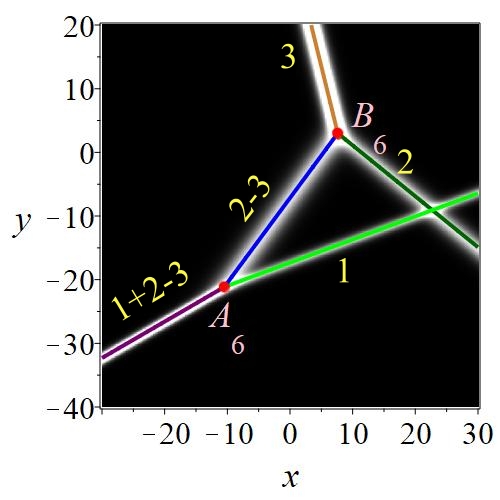}}
\vspace{-0.6\baselineskip}
\caption{The density plots of the weak 2-resonance 3-soliton with 
$k_1=1,\,k_2=-1,\,k_3=-2,\,p_1 = -\frac{k_1 \left( k_1 k_3 - k_3^2 - p_3 \right)}{k_3}, \, p_2 = \frac{k_2 \left( k_2 k_3 - k_3^2 +p_3 \right)}{k_3},\,p_3=-\frac{1}{2}$. 
The lines are the trajectories of the arms and stem structures, and the points are the endpoints of the variable length stem structures.
}\label{fig2-7}
\end{figure}

Proposition~\ref{prop2.2} reveals that the weak 2-resonant 3-soliton solution exhibits a characteristic four-arm configuration accompanied by a stem structure as \( t \to \pm\infty \), similar to the structure observed in the strong 2-resonant case. 
In both cases, the soliton arms \( S_1 \) and \( S_2 \) experience a phase shift before and after their interaction, which can be attributed to the influence of the coefficient \( a_{12} \). 
However, a fundamental difference lies in the nature of the interactions involving \( S_3 \). 
Unlike in the strong 2-resonant configuration, the coefficients \( a_{13} \) and \( a_{23} \) in this case satisfy the weak resonance conditions given in Eq.~\eqref{3f2r5}. 
Consequently, weak resonances occur between \( S_1 \) and \( S_3 \), and between \( S_2 \) and \( S_3 \), leading to the emergence of two intermediate solitons, \( S_{1-3} \) and \( S_{2-3} \). 
A further resonance process involving these intermediate states produces the composite soliton \( S_{1+2-3} \).

The evolution of the soliton interaction, including the reconnection of the four soliton arms and the transient appearance and disappearance of stem structures, is illustrated in Fig.~\ref{fig2-7}. 
As \( t \to -\infty \), the stem structure \( S_{1-3} \) connects two pairs of V-shaped solitons: \( (S_1, S_3) \) and \( (S_2, S_{1+2-3}) \). 
As time progresses, the stem \( S_{1-3} \) gradually shortens and eventually vanishes near \( t = 0 \), at which point all four soliton arms (\( S_1, S_2, S_3, S_{1+2-3} \)) merge at a common location. 
This results in a reconnection of soliton pairs, forming \( (S_1, S_{1+2-3}) \) and \( (S_2, S_3) \). 
As time continues to evolve (i.e. \( t \to +\infty \)), a new stem structure \( S_{2-3} \) emerges and gradually increases in length, ultimately connecting the newly formed V-shaped soliton pairs.

The amplitudes and propagation velocities of the soliton arms and stem structures are presented in Table~\ref{tab:t1}, and the equations describing their trajectories are as follows:
\vspace{-0.4\baselineskip}
\begin{flalign}\label{l02}
\boldsymbol{l_j:}\,\xi_j=0,\quad \boldsymbol{\widehat{l_j}:}\,\xi_j+\ln a_{12}=0,\quad \boldsymbol{\widehat{l_{1+2-3}}:}\,\xi_1+\xi_2-\xi_3+\ln a_{12}=0.
\vspace{-0.6\baselineskip}
\end{flalign}

Geometric analysis shows that the trajectories of \( S_1 \), \( S_3 \), and \( S_{1-3} \) intersect at a single point as \( t \to -\infty \), and similarly, the trajectories of \( S_2 \), \( S_{1+2-3} \), and \( S_{1-3} \) also intersect at a common point. 
At \( t \to +\infty \), analogous intersections occur among the trajectories of \( S_2 \), \( S_3 \), and \( S_{2-3} \), as well as \( S_1 \), \( S_{1+2-3} \), and \( S_{2-3} \). 
These consistent intersection points further confirm the analytical predictions for the stem locations given in Eq.~\eqref{3s2rstem01}. 
Figure~\ref{fig2-7} presents the time-evolving trajectories of the solitons, superimposed on a background density plot of the solution \( u \). 
By solving for the intersection points of these trajectories, one can precisely identify the endpoints of the variable-length stem structures as follows:
\vspace{-0.5\baselineskip}
\begin{flalign}\label{endpoints2}
\begin{split}
A_5\,&\Bigg(\bigg(k_3^2-2p_3+\frac{4k_{1}k_{3}p_{3} - 3p_{3}^{2}}{k_{3}^{2}}\bigg)t,\,\bigg(\frac{6p_3}{k_3}-4k_1+2k_3\bigg)t\Bigg),\,\\
B_5\,&\Bigg(\frac{(p_3-k_1k_3)\ln a_{12}}{k_2k_3(k_1+k_2-k_3)}+\bigg(k_3^2-4k_1k_3+4k_1k_2+2p_3+\frac{4k_1p_3-4k_2p_3}{k_3}-\frac{3p_3^2}{k_3^2}\bigg)t,\\
&-\frac{\ln a_{12}}{k_2(k_1+k_2-k_3)} +\bigg(\frac{6p_3}{k_3}-4k_1+4k_2-2k_3\bigg)t \Bigg),\,\\
A_6\,&\Bigg(\bigg(k_3^2+2p_3-\frac{4k_{2}k_{3}p_{3} + 3p_{3}^{2}}{k_{3}^{2}}\bigg)t,\,\bigg(\frac{6p_3}{k_3}+4k_2-2k_3\bigg)t\Bigg),\,\\
B_6\,&\Bigg(-\frac{(p_3+k_2k_3)\ln a_{12}}{k_1k_3(k_1+k_2-k_3)}+\bigg(k_3^2-4k_2k_3+4k_1k_2-2p_3+\frac{4k_1p_3-4k_2p_3}{k_3}-\frac{3p_3^2}{k_3^2}\bigg)t,\\
&\frac{\ln a_{12}}{k_1(k_1+k_2-k_3)}+\bigg(\frac{6p_3}{k_3}-4k_1+4k_2+2k_3\bigg)t \Bigg).
\vspace{-0.5\baselineskip}
\end{split}
\end{flalign}
In this context, point \( A_5 \) corresponds to the intersection of the trajectories \( l_1 \) and \( l_3 \), while point \( B_5 \) denotes the intersection of the trajectories \( \widehat{l_2} \) and \( \widehat{l_{1+2-3}} \). 
Similarly, point \( A_6 \) marks the intersection of \( l_2 \) and \( l_3 \), and point \( B_6 \) represents the intersection of \( \widehat{l_1} \) and \( \widehat{l_{1+2-3}} \). 
These key points are also illustrated in Fig.~\ref{fig2-7}. 
Based on these intersection points, the lengths of the variable-length stem trajectories can be determined as follows:
\vspace{-0.5\baselineskip}
\begin{flalign}\label{3length-2}
\begin{split}
&|A_5B_5|=\sqrt{k_1^2k_3^2-2k_1k_3p_3+k_3^2+p_3^2}\left|\frac{\ln a_{12}}{k_2k_3(k_1+k_2-k_3)}-\frac{4(k_2-k_3)t}{k_3} \right|,\,t\ll 0,\\
&|A_6B_6|=\sqrt{k_2^2k_3^2+2k_2k_3p_3+k_3^2+p_3^2}\left|\frac{\ln a_{12}}{k_1k_3(k_1+k_2-k_3)}-\frac{4(k_1-k_3)t}{k_3} \right|,\,t\gg 0.
\vspace{-0.5\baselineskip}
\end{split}
\end{flalign}
Due to the similarly complex dynamics near $t=0$, as discussed in the previous section, Eq.\ \ref{3length-2} is limited to the regime $|t|\ll 0$.

To further characterize the variable-length stem structures, we examine their amplitude profiles. 
Specifically, the cross-sectional curves of the 3-soliton solution \eqref{3f2r5} along the trajectories \( l_{1-3} \) and \( l_{2-3} \), as shown in Fig.~\ref{fig2-7}, are given by:
\begin{flalign}\label{cross3s02}
\begin{split}
&u|_{l_{1-3}}=\frac{2a_{12}k_1^2\e^{\theta_5+2\theta_6}+4a_{12}k_2^2\e^{2\theta_5+\theta_6}+(2a_{12}(k_1+k_2)^2+4(k_1-k_2)^2)\e^{\theta_5+\theta_6}+4k_1^2\e^{\theta_5}
+2k_2^2\e^{\theta_6}}{(1+2\e^{\theta_5}+\e^{\theta_6}+a_{12}\e^{\theta_5+\theta_6})^2},\\
&u|_{l_{2-3}}=\frac{4a_{12}k_1^2\e^{\theta_7+2\theta_8}+2a_{12}k_2^2\e^{2\theta_7+\theta_8}+(2a_{12}(k_1+k_2)^2+4(k_1-k_2)^2)\e^{\theta_7+\theta_8}+2k_1^2\e^{\theta_7}
+4k_2^2\e^{\theta_8}}{(1+\e^{\theta_7}+2\e^{\theta_8}+a_{12}\e^{\theta_7+\theta_8})^2},
\end{split}
\end{flalign}
where
\begin{flalign*}
&\theta_5=\frac{k_1(k_3^2x-(k_3^4+4k_1k_3p_3-2k_3^2p_3-3p_3^2)t)}{k_1k_3-p_3},\,\\
&\theta_6=\frac{4k_2(k_1+k_2-k_3)(k_{3}^{2}x-(k_{3}^{4}-4k_{1}k_{3}^{3}+(4k_{1}k_{2} + 2p_{3})k_{3}^{2}+4p_{3}(k_{1} - k_{2})k_{3}-3p_{3}^{2})t) }{k_3(k_1k_3-p_3)},\\
&\theta_7=\frac{4k_1(k_1+k_2-k_3)(k_{3}^{2}x-(k_{3}^{4}-4k_{2}k_{3}^{3}+(4k_{1}k_{2} - 2p_{3})k_{3}^{2}+4p_{3}(k_{1} - k_{2})k_{3}-3p_{3}^{2})t) }{k_3(k_2k_3+p_3)},\\
&\theta_8=\frac{k_2(k_3^2x-(k_3^4-4k_2k_3p_3+2k_3^2p_3-3p_3^2)t)}{k_2k_3+p_3}.
\end{flalign*}

\begin{figure}[h!tb]
\centering
\subfigure[$t=-2$]{\includegraphics[height=3cm,width=3cm]{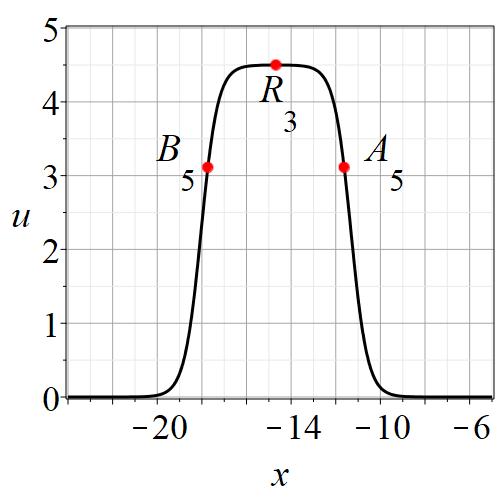}}
\subfigure[$t=2$]{\includegraphics[height=3cm,width=3cm]{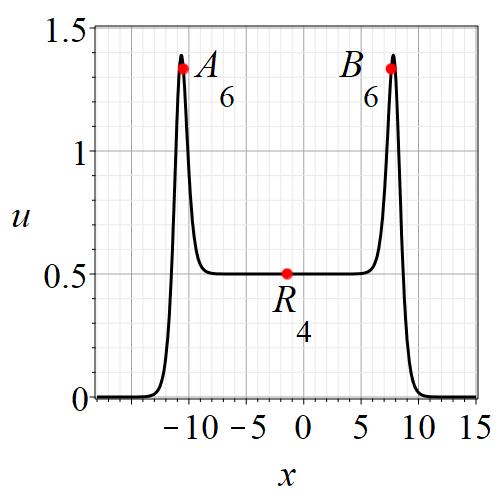}}
\subfigure[]{\includegraphics[height=3cm,width=3cm]{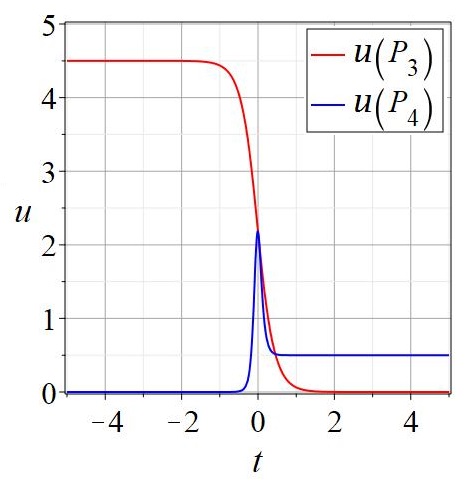}}
\subfigure[$t=-2$]{\includegraphics[height=3cm,width=3cm]{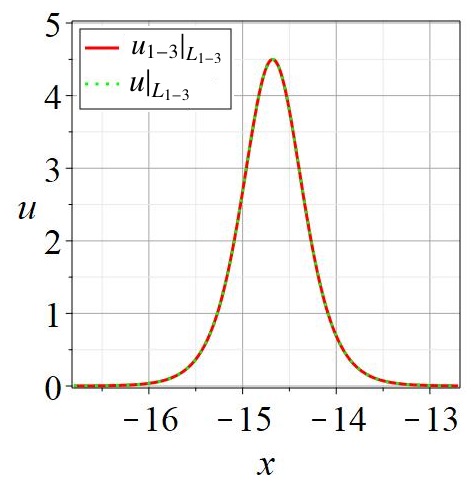}}
\subfigure[$t=2$]{\includegraphics[height=3cm,width=3cm]{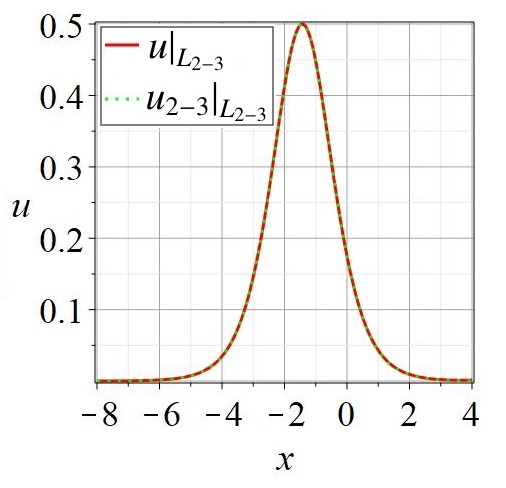}}
\vspace{-0.6\baselineskip}
\caption{Parameters: $k_1=1,\,k_2=-1,\,k_3=-2,\,p_1 = -\frac{k_1 \left( k_1 k_3 - k_3^2 - p_3 \right)}{k_3}, \, p_2 = \frac{k_2 \left( k_2 k_3 - k_3^2 +p_3 \right)}{k_3},\,p_3=-\frac{1}{2}$. 
(a) The cross-sectional curves $u|_{l_{1-3}}$ given by \eqref{cross3s02}; 
(b) The cross-sectional curves $u|_{l_{2-3}}$ given by \eqref{cross3s02};
(c) The amplitude evolution curves \eqref{up02}; 
(d) The cross-sectional curves \( u|_{L_{1-3}^{(1)}} \) and \( u_{1-3}|_{L_{1-3}^{(1)}} \); 
(e) The cross-sectional curves \( u|_{L_{2-3}^{(1)}} \) and \( u_{2-3}|_{L_{2-3}^{(1)}} \).
}\label{fig2-8}
\end{figure}

Due to the complexity of the calculations, the extreme points of the cross-sectional curve \eqref{cross3s02} cannot be obtained analytically. 
Instead, for given parameters \((k_j, p_j)\) and $t$, numerical methods are employed to determine the extreme points and corresponding amplitudes along the stem structure.
For instance, with the parameter set used in Fig.~\ref{fig2-7}~(a), the extreme point is located at \((-14.679, 8.928)\) with an extreme value \(4.499\approx \frac{(k_1 - k_3)^2}{2} = \frac{9}{2}\). 
This is indicated by the red point \(R_3\) in Fig.~\ref{fig2-8}~(a).
Similarly, for the parameters shown in Fig.~\ref{fig2-7}~(c), the extreme point is found at \((-1.429, -9.072)\) with an extreme value \(0.500\approx \frac{(k_2 - k_3)^2}{2} = \frac{1}{2}\). 
This corresponds to the red point \(R_4\) in Fig.~\ref{fig2-8}~(b).
Furthermore, the amplitudes at the midpoints of segments \(A_5 B_5\) and \(A_6 B_6\), denoted by \(P_3\) and \(P_4\) respectively, are expressed as follows:
\begin{flalign} \label{up02}
\begin{split}
u(P_3)={\frac {30\,\sqrt {3}{{\rm e}^{4\,t}}+6\,\sqrt {3}{{\rm e}^{12\,t}}+30\,{{\rm e}^{8\,t}}+54}{ \left( 3\,{{\rm e}^{4\,t}}+ \left( {{\rm e}^{8
\,t}}+2 \right) \sqrt {3} \right) ^{2}}},\,
u(P_4)=6\,{\frac { \left( 5\,\sqrt {3}{{\rm e}^{24\,t}}+{{\rm e}^{36\,t}}+\sqrt {3}+13\,{{\rm e}^{12\,t}} \right) {{\rm e}^{12\,t}}}{ \left( 2\,
\sqrt {3}{{\rm e}^{24\,t}}+\sqrt {3}+3\,{{\rm e}^{12\,t}} \right) ^{2}}}.
\end{split}
\end{flalign}
The amplitude evolution of \( u(P_3) \) and \( u(P_4) \) is depicted in Fig.~\ref{fig2-8}~(c). 
As observed, the amplitudes given by \eqref{up02} exhibit strong fluctuations and complexity near \( t = 0 \), but gradually stabilize as \( |t| \gg 0 \). 
This behavior arises because, around \( t = 0 \), the four soliton arms are in close proximity and interact strongly, rendering the stem structures indistinct. 
Consequently, analyzing the stem properties in this regime offers limited insight. 
Moreover, the limiting behavior of the amplitudes confirms the validity of using \( u(P_3) \) and \( u(P_4) \) as approximations for the amplitudes of stem structures \( S_{1-3} \) and \( S_{2-3} \), respectively. 
Specifically,
\[\lim_{t \to -\infty} u(P_3) = \frac{9}{2} = \frac{(k_1 - k_3)^2}{2}, \quad \lim_{t \to +\infty} u(P_4) = \frac{1}{2} = \frac{(k_2 - k_3)^2}{2}.\]

To further examine this correspondence, we consider vertical planes orthogonal to the stem directions and passing through the points \( P_3 \) and \( P_4 \). 
The plane perpendicular to \( l_{1-3} \) and passing through \( P_3 \) is defined by
\[L_{1-3}^{(1)}: \quad (p_1 - p_3)x - (k_1 - k_3)y - \left[(p_1 - p_3)x_{P_3} - (k_1 - k_3)y_{P_3}\right] = 0,\]
where \( (x_{P_3}, y_{P_3}) \) are the coordinates of point \( P_3 \). 
Fig.~\ref{fig2-8} (d) displays the intersection curves of \( u \) and \( u_{1-3} \) with this plane, denoted by \( u|_{L_{1-3}^{(1)}} \) and \( u_{1-3}|_{L_{1-3}^{(1)}} \), respectively.

Similarly, the plane orthogonal to \( l_{2-3} \) and passing through \( P_4 \) is given by
\[L_{2-3}^{(1)}: \quad (p_2 - p_3)x - (k_2 - k_3)y - \left[(p_2 - p_3)x_{P_4} - (k_2 - k_3)y_{P_4}\right] = 0,\]
with \( (x_{P_4}, y_{P_4}) \) denoting the coordinates of \( P_4 \). 
The corresponding intersection curves \( u|_{L_{2-3}^{(1)}} \) and \( u_{2-3}|_{L_{2-3}^{(1)}} \) are shown in Fig.~\ref{fig2-8} (e).

From Figs.~\ref{fig2-8} (d) and (e), it is evident that the profiles of \( u \) and its approximations \( u_{1-3} \) and \( u_{2-3} \) nearly coincide. 
This further supports the validity of employing \( u_{1-3} \) and \( u_{2-3} \) as asymptotic approximations to the full solution \( u \) as \( t \to -\infty \) and \( t \to +\infty \), respectively.

\subsection{Mixed 2-resonant case: $a_{13} \to +\infty$ and $a_{23} \to 0$}
To derive the mixed (strong-weak) 2-resonant 3-soliton solution, we take the limits $a_{13} \to +\infty$ and $a_{23} \to 0$. 
As in Subsection~2.1, this is achieved via a transformation. 
Specifically, we perform the substitution $\xi_3 \to \xi_3 - \ln a_{13}$ in Eq.~\eqref{3f}, and then take the limits $a_{13} \to +\infty$ and $a_{23} \to 0$. 
As a result, we obtain the strong-weak 2-resonant 3-soliton solution in the form:
\begin{flalign}\label{3f2r6}
u=2(\ln f)_{xx},\,f=1+\exp\xi_1+\exp\xi_2+a_{12}\exp(\xi_1+\xi_2)+\exp(\xi_1+\xi_3).
\end{flalign}

To ensure the asymptotic behavior $a_{13} \to +\infty$, $a_{23} \to 0$, and $0 < a_{12} < +\infty$, the parameters must satisfy one of the following two conditions:
\[p_1 = \frac{k_1 (k_1 k_3 + k_3^2 + p_3)}{k_3}, \quad p_2 = \frac{k_2 (k_2 k_3 - k_3^2 + p_3)}{k_3},\]
or
\[p_1 = -\frac{k_1 (k_1 k_3 + k_3^2 - p_3)}{k_3}, \quad p_2 = -\frac{k_2 (k_2 k_3 - k_3^2 - p_3)}{k_3}.\]
Without loss of generality, we consider only the first case in this subsection. 
Under this condition, the interaction coefficient $a_{12}$ is given by $a_{12} = -\frac{k_3 (k_1 - k_2 + k_3)}{(k_1 + k_3)(k_2 - k_3)}$.
This leads to the following proposition:
\begin{prop}\label{prop2.3}
The asymptotic forms of the strong-weak 2-resonant 3-soliton \eqref{3f2r6} with 
$p_1 = \frac{k_1 ( k_1 k_3 + k_3^2 + p_3)}{k_3}$ and $p_2 = \frac{k_2 ( k_2 k_3 -k_3^2 +p_3)}{k_3}$ are as following:

Before collision ($t\to-\infty$):
\begin{flalign}\label{3s2rasy05}
\begin{split}
y\to -\infty,\quad &S_1:\quad u\sim \widehat{u_1},\quad S_2:\quad u\sim \widehat{u_2},\\
y\to +\infty,\quad &S_3:\quad u\sim u_3,\quad S_{1-2+3}:\,\,u\sim u_{1-2+3}.
\end{split}
\end{flalign}

After collision ($t\to+\infty$):
\begin{flalign}\label{3s2rasy06}
\begin{split}
y\to -\infty,\quad &S_1:\quad u\sim u_1,\quad S_2:\quad u\sim u_2,\\
y\to +\infty,\quad &S_3:\quad u\sim u_3,\quad S_{1-2+3}:\,\,u\sim u_{1-2+3}.
\end{split}
\end{flalign}

The stem structures:
\begin{flalign}\label{3s2rstem03}
\begin{split}
t\to-\infty,\quad S_{2-3}:\, u\sim \widehat{u_{2-3}};\quad t\to+\infty,\quad S_{1+3}:\, u\sim u_{1+3}.
\end{split}
\end{flalign}

Here, the corresponding formulas are presented in Eqs.\ \eqref{uj01}, \eqref{uj02} and
\vspace{-0.2\baselineskip}
\begin{flalign}\label{uj03}
\begin{split}
\widehat{u_{i- j}}=\frac{(k_i- k_j)^2}{2}\sech^2\frac{\xi_i- \xi_j+\ln a_{12}}{2},\,u_{1-2+3}= \frac{(k_1-k_2+k_3)^2}{2}\sech^2\frac{\xi_1-\xi_2+\xi_3}{2}.
\end{split}
\end{flalign}
\end{prop}

\begin{figure}[h!tb]
\centering
\subfigure[$t=-8$]{\includegraphics[height=4cm,width=5cm]{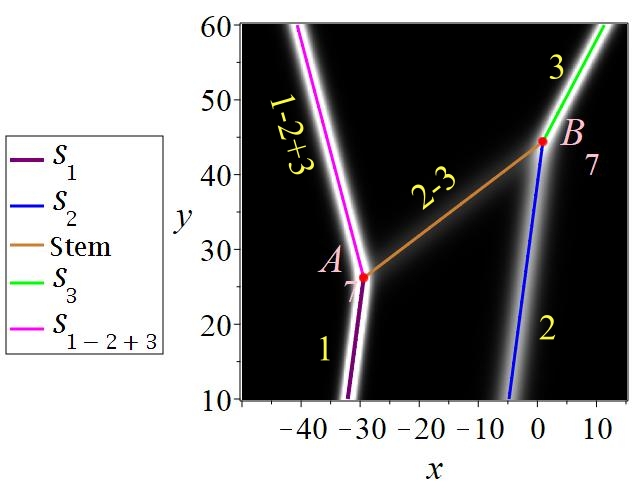}}
\subfigure[$t=0$]{\includegraphics[height=4cm,width=4cm]{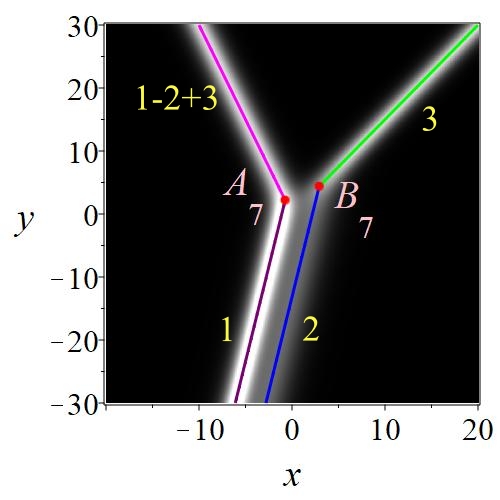}}
\subfigure[$t=10$]{\includegraphics[height=4cm,width=4cm]{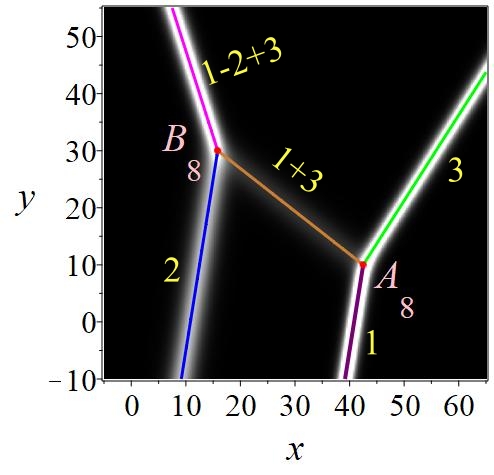}}
\vspace{-0.6\baselineskip}
\caption{The density plots of the strong-weak 2-resonant 3-soliton with 
$k_1=2,\,k_2=-1,\,k_3=-\frac{3}{2},\,p_1 = \frac{k_1 \left( k_1 k_3 + k_3^2 + p_3 \right)}{k_3}, \, p_2 = \frac{k_2 \left( k_2 k_3 -k_3^2 +p_3 \right)}{k_3},\,p_3=1$. 
The lines are the trajectories of the arms and stem structures, and the points are the endpoints of the variable-length stem structures.
}\label{fig2-9}
\end{figure}

The reconnection of four soliton arms, along with the disappearance and subsequent fission of the stem structure, is illustrated in Fig.~\ref{fig2-9}. 
As $t \to -\infty$, the stem structure $S_{2-3}$ connects two pairs of V-shaped solitons: $(S_1,\, S_{1-2+3})$ and $(S_2,\, S_3)$. 
As time progresses, the length of the stem $S_{2-3}$ gradually decreases and vanishes near $t = 0$. 
At this moment, the four soliton arms ($S_1$, $S_2$, $S_3$, and $S_{1-2+3}$) merge at a common location, and the two initial V-shaped soliton pairs effectively transform into the same configurations, $S_1$ with $S_{1-2+3}$ and $S_2$ with $S_3$. 
As time continues (i.e. $t \to +\infty$), a new stem structure $S_{1+3}$ emerges and gradually elongates, thereby reconnecting the two V-shaped soliton pairs.
The amplitudes and velocities of the four soliton arms and the two stem structures are listed in Table~\ref{tab:t1}, and their corresponding trajectories are given by
\vspace{-0.2\baselineskip}
\begin{flalign}\label{l03}
\boldsymbol{\widehat{l_{i- j}}:}\, \xi_i- \xi_j+\ln a_{12}=0,\,\boldsymbol{l_{1-2+3}}:\,\xi_1-\xi_2+\xi_3=0.
\end{flalign}
Fig.~\ref{fig2-9} presents the trajectories of the solution $u$ at various time instances, where the background represents a density plot. 
In this figure, the soliton arms do not exhibit an X-shaped intersection because the chosen parameters result in the arms $S_{1-2+3}$ and $S_2$ being nearly parallel. 
By selecting more general parameters, X-shaped crossings---similar to those shown in Figs.~\ref{fig2-6} (c) and \ref{fig2-7} (c)---can also occur. 
However, these differences arise solely from the choice of parameters and do not reflect any essential change in the nature of the soliton solution. 
Therefore, we do not pursue a separate analysis here.

\begin{figure}[h!tb]
\centering
\subfigure[$t=-8$]{\includegraphics[height=3cm,width=3cm]{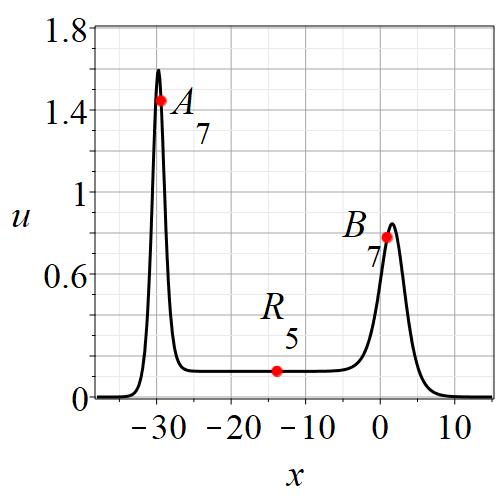}}
\subfigure[$t=10$]{\includegraphics[height=3cm,width=3cm]{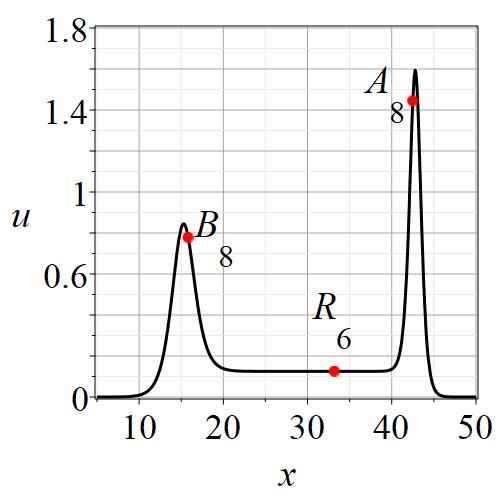}}
\subfigure[]{\includegraphics[height=3cm,width=3cm]{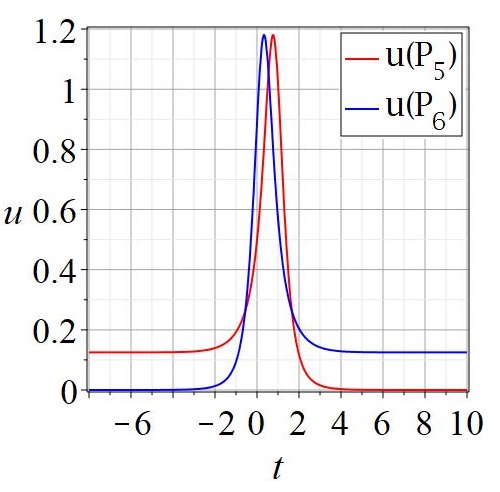}}
\subfigure[$t=-8$]{\includegraphics[height=3cm,width=3.2cm]{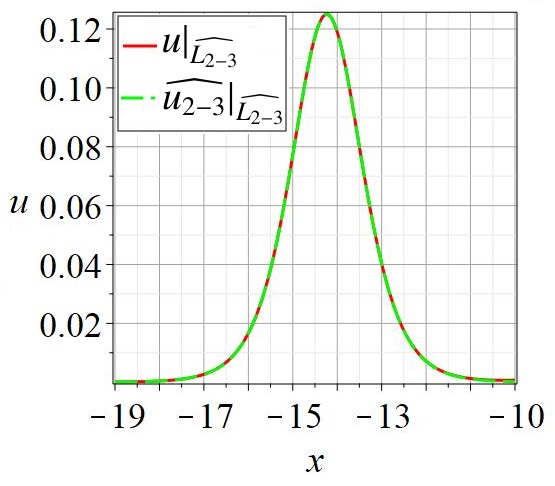}}
\subfigure[$t=10$]{\includegraphics[height=3cm,width=3.2cm]{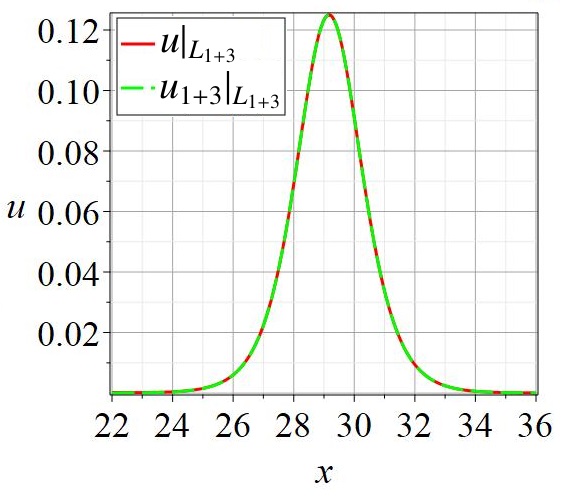}}
\vspace{-0.6\baselineskip}
\caption{Parameters: $k_1=2,\,k_2=-1,\,k_3=-\frac{3}{2},\,p_1 = \frac{k_1 \left( k_1 k_3 + k_3^2 + p_3 \right)}{k_3}, \, p_2 = \frac{k_2 \left( k_2 k_3 -k_3^2 +p_3 \right)}{k_3},\,p_3=1$. 
(a) The cross-sectional curves $u|_{l_{1+3}}$ given by \eqref{cross3s03}; 
(b) The cross-sectional curves $u|_{\widehat{l_{2-3}}}$ given by \eqref{cross3s03}; 
(c) The amplitude evolution curves \eqref{up03}; 
(d) The cross-sectional curves $u|_{L_{2-3}^{(2)}}$ and $\widehat{u_{2-3}}|_{L_{2-3}^{(2)}}$; 
(e) The cross-sectional curves $u|_{L_{1+3}}$ and $u_{1+3}|_{L_{1+3}}$.
}\label{fig2-10}
\end{figure}

By computing the intersection points of these trajectories, the endpoints of the variable-length stem structures can be determined as follows:
\vspace{-0.2\baselineskip}\begin{flalign}\label{endpoints3}
\begin{split}
A_7\,&\Bigg(-\frac{(k_1k_3+k_2k_3+p_3)\ln a_{12}}{k_1k_3(k_2-k_3)}+\bigg(k_3^2-4k_1k_2-4k_2k_3-2p_3-\frac{4p_3(k_1+k_2)}{k_3}-\frac{3p_3^2}{k_3^2}\bigg)t,\\
&\frac{\ln a_{12}}{k_1(k_2-k_3)} +\bigg(\frac{6p_3}{k_3}+4k_1+4k_2+2k_3\bigg)t \Bigg),\\
B_7\,&\Bigg(\frac{p_3\ln a_{12}}{k_2k_3(k_2-k_3)}+\bigg(k_3^2+2p_3-\frac{4k_2p_3}{k_3}-\frac{3p_3^2}{k_3^2}\bigg)t,\,\frac{-\ln a_{12}}{k_2(k_2-k_3)} +\bigg(\frac{6p_3}{k_3}+4k_2-2k_3\bigg)t \Bigg),\\
A_8\,&\Bigg(\bigg(k_3^2-2p_3-\frac{4p_3k_1}{k_3}-\frac{3p_3^2}{k_3^2}\bigg)t,\,\bigg(\frac{6p_3}{k_3}+4k_1+2k_3\bigg)t \Bigg),\\
B_8\,&\Bigg(\bigg(k_3^2-4k_1k_2+4k_1k_3+2p_3-\frac{4k_1p_3+4p_3k_2}{k_3}-\frac{3p_3^2}{k_3^2}\bigg)t,\,\bigg(\frac{6p_3}{k_3}+4k_1+4k_2-2k_3\bigg)t \Bigg).
\end{split}
\end{flalign}
In this context, point $A_7$ corresponds to the intersection of the lines $\widehat{l_1}$ and $l_{1-2+3}$, while point $B_7$ denotes the intersection of $\widehat{l_2}$ and $l_3$. 
Similarly, point $A_8$ marks the intersection of $l_1$ and $l_{1-2+3}$, and point $B_8$ represents the intersection of $l_2$ and $l_3$. 
These intersection points are also depicted in Fig.~\ref{fig2-10}. 
Based on these, the lengths of the trajectories of the variable-length stem structures can be calculated as follows:
\vspace{-0.2\baselineskip}
\begin{flalign}\label{3length-3}
\begin{split}
&|A_7B_7|=\sqrt{k_3^2+(k_2k_3+p_3)^2}\left|\frac{(k_1+k_2)\ln a_{12}}{k_1k_2k_3(k_2-k_3)}+\frac{4(k_1+k_3)t}{k_3} \right|,\,t\ll 0,\\
&|A_8B_8|=4|t(k_2-k_3)|\sqrt{k_3^2+(k_1k_3+p_3)^2},\,t\gg 0.
\end{split}
\end{flalign}
As in Ssubsections~2.1-2.2, due to the complex soliton dynamics near $t = 0$, Eq.~\eqref{3length-3} is valid only for $|t|\ll 0$.

Now we will analyze the amplitudes of variable-length stem structures next. 
The cross-sectional curves of 3-soloton \eqref{3f} with Eqs.\ \eqref{3f2r6} along $l_{1+3}$ and $\widehat{l_{2-3}}$ shown in Fig.\ \ref{fig2-9} are expressed as,
\vspace{-0.2\baselineskip}
\begin{flalign}\label{cross3s03}
\begin{split}
&u|_{l_{1+3}}=\frac{a_{12}k_1^2\e^{\theta_9+2\theta_{10}}+2a_{12}k_2^2\e^{2\theta_9+\theta_{10}}+(4a_{12}(k_1+k_2)^2+2(k_1-k_2)^2)\e^{\theta_9+\theta_{10}}
+4k_1^2\e^{\theta_9}+4k_2^2\e^{\theta_{10}}}{(2+\e^{\theta_9}+\e^{\theta_{10}}+a_{12}\e^{\theta_9+\theta_{10}})^2},\\
&u|_{\widehat{l_{2-3}}}=\frac{4a_{12}k_1^2\e^{\theta_{11}+2\theta_{12}}+4a_{12}k_2^2\e^{2\theta_{11}+\theta_{12}}+(4a_{12}(k_1+k_2)^2+2(k_1-k_2)^2)\e^{\theta_{11}
+\theta_{12}}+2k_1^2\e^{\theta_{11}}+2k_2^2\e^{\theta_{12}}}{(1+\e^{\theta_{11}}+\e^{\theta_{12}}+2a_{12}\e^{\theta_{11}+\theta_{12}})^2},
\end{split}
\end{flalign}
where
\vspace{-0.2\baselineskip}
\begin{flalign*}
&\theta_9=\frac{(k_1p_3-k_3p_1)x}{p_1+p_3}+\bigg(\frac{k_3^3p_1-k_1^3p_3}{p_1+p_3}+\frac{3k_1p_1p_3^2-3k_3p_1^2p_3}{k_1k_3(p_1+p_3)}\bigg)t,\\
&\theta_{10}=\bigg(k_2-\frac{k_1p_2+k_3p_2}{p_1+p_3}\bigg)x+\bigg(\frac{k_1^3p_2+k_3^3p_2}{p_1+p_3}+\frac{3k_1p_2p_3^2+3k_3p_1^2p_2}{k_1k_3(p_1+p_3)}
-k_2^3-\frac{3p_2^3}{k_2}\bigg)t,\\
&\theta_{11}=\bigg(k_1-\frac{k_2p_1-k_3p_1}{p_2-p_3}\bigg)x+\bigg(\frac{k_2^3p_1-k_3^3p_1}{p_2-p_3}-\frac{3k_2p_1p_3^2-3k_3p_2^2p_1}{k_2k_3(p_2-p_3)}
-k_2^3-\frac{3p_1^3}{k_1}\bigg)t-\frac{p_1\ln a_{12}}{p_2-p_3},\\
&\theta_{12}=\frac{(k_3p_2-k_2p_3)x}{p_2-p_3}+\bigg(\frac{k_2^3p_3-k_3^3p_2}{p_2-p_3}-\frac{3k_2p_2p_3^2-3k_3p_2^2p_3}{k_2k_3(p_2-p_3)}\bigg)t
-\frac{p_2\ln a_{12}}{p_2-p_3}.
\end{flalign*}

Due to the complexity of the calculation, the extreme points of the cross-sectional curve in Eq.~\eqref{cross3s03} cannot be obtained analytically. 
However, for given parameters $(k_j,\, p_j)$, and $t$, numerical methods can be used to determine the locations and amplitudes of the extreme points between the two endpoints of the stem structure. 
For instance, when the parameters are the same as those used in Fig.~\ref{fig2-9}~(a), the numerical extreme point is $(-13.832,\ 35.538)$, with an amplitude of $0.125 \approx \frac{(k_2 - k_3)^2}{2} = \frac{1}{8}$, as indicated by the red point $R_5$ in Fig.~\ref{fig2-10}~(a). 
Similarly, when the parameters correspond to those in Fig.~\ref{fig2-9}~(c), the extreme point is located at $(33.200,\ 16.975)$, with an amplitude of $0.125 \approx \frac{(k_1 + k_3)^2}{2} = \frac{1}{8}$, as shown by the red point $R_6$ in Fig.~\ref{fig2-10}~(b). 
If we examine the amplitude at the midpoints of the segments $A_7B_7$ and $A_8B_8$, denoted by $P_5$ and $P_6$ respectively, they can be expressed as:
\vspace{-0.2\baselineskip}
\begin{flalign} 
\begin{split}\label{up03}
&u(P_5)=\frac{3 \e^{-\frac{3t}{2}} \left(729\e^{-\frac{15t}{2}} + 1053\e^{-6t}\sqrt{3} + 675\e^{-\frac{9t}{2}} + 243\e^{-3t}\sqrt{3} + 144\e^{-\frac{3t}{2}} 
+ 4\sqrt{3}\right)}{2\left(\sqrt{3}\e^{-\frac{3t}{2}} + 54\sqrt{3}\e^{-\frac{9t}{2}} + 9\e^{-3t} + 3\right)^2},\\
&u(P_6)=\frac{36\e^{-\frac{15t}{2}} + 144\e^{-6t} + 81\e^{-\frac{9t}{2}} + 25\e^{-3t} + 13\e^{-\frac{3t}{2}} + 1}{2\left(9\e^{-\frac{9t}{2}} + \e^{-3t} 
+ \e^{-\frac{3t}{2}} + 2\right)^2}.
\end{split}
\end{flalign}
The amplitude trend plots of $u(P_5)$ and $u(P_6)$ are shown in Fig.~\ref{fig2-10} (c). 
As observed from the figure, the amplitudes given by Eq.~\eqref{up03} exhibit strong and complex fluctuations near $t = 0$, but gradually stabilize as $|t| \gg 0$. 
This behavior arises because, near $t = 0$, the four soliton arms are in close proximity and interact strongly, making the stem structure difficult to distinguish. 
Hence, analyzing the stem properties around $t = 0$ offers limited insight. 
Moreover, since
\vspace{-0.2\baselineskip}
\[\lim_{t \to -\infty} u(P_5) = \frac{1}{8} = \frac{(k_2 - k_3)^2}{2}, \quad 
\lim_{t \to +\infty} u(P_6) = \frac{1}{8} = \frac{(k_1 + k_3)^2}{2},\]
it is reasonable to use $u(P_5)$ and $u(P_6)$ as approximations for the amplitudes of the stem structures $S_{2-3}$ and $S_{1+3}$, respectively.

In addition, consider the vertical plane passing through point $P_5$ and perpendicular to the line $\widehat{l_{2-3}}$, which is defined by:
\vspace{-0.2\baselineskip}
\[L_{2-3}^{(2)}: \quad (p_2 - p_3)x - (k_2 - k_3)y - \left[ (p_2 - p_3)x_{P_5} - (k_2 - k_3)y_{P_5} \right] = 0,\]
where $x_{P_5}$ and $y_{P_5}$ are the coordinates of point $P_5$. 
Fig.~\ref{fig2-10} (d) displays the intersection curves of $u$ and $\widehat{u}_{2-3}$ with the plane $\widehat{L_{2-3}}$, 
denoted by $u|_{L_{2-3}^{(2)}}$ and $\widehat{u}_{2-3}|_{L_{2-3}^{(2)}}$, respectively.

Similarly, the vertical plane passing through point $P_6$ and perpendicular to $l_{1+3}$ is given by:
\vspace{-0.2\baselineskip}
\[L_{1+3}: \quad (p_1 + p_3)x - (k_1 + k_3)y - \left[ (p_1 + p_3)x_{P_6} - (k_1 + k_3)y_{P_6} \right] = 0,\]
where $x_{P_6}$ and $y_{P_6}$ denote the coordinates of point $P_6$. 
Fig.~\ref{fig2-10} (e) shows the intersection curves of $u$ and $u_{1+3}$ with the plane $L_{1+3}$, 
denoted by $u|_{L_{1+3}}$ and $u_{1+3}|_{L_{1+3}}$, respectively.

As illustrated in Fig.~\ref{fig2-10} (d) and (e), these curves nearly coincide, further supporting the validity of using $\widehat{u}_{2-3}$ and $u_{1+3}$ as approximations to $u$ for $t \to -\infty$ and $t \to +\infty$, respectively.

\begin{table}[ht]
\centering
\caption{\normalsize Physical quantities of the soliton arms}
\label{tab:t1}
\begin{tabular}{cccccc}
\Xhline{1pt}
Arms & Trajectory & Velocity on $(x,\,y)$-direction& Amplitude & Formulas \\
\hline
\multirow{2}{*}{$S_{j}$} & $l_j$ & \multirow{2}{*}{$\left(k_j^2+\frac{3p_j^2}{k_j}, \frac{k_j^4+3p_j^2}{k_jp_j}\right)$} & \multirow{2}{*}{$\frac{k_j^2}{2}$} & $u_{j}$ \\
&$\widehat{l_j}$ & & & $\widehat{u_{j}}$ \\
\hline
\multirow{2}{*}{$S_{i+j}$} & $l_{i+j}$ & \multirow{2}{*}{$\left(k_i^2-k_ik_j+k_j^2+\frac{3p_i^2k_j+3p_j^2k_i}{k_ik_j(k_i+k_j)}, \frac{k_i^3+k_j^3}{p_i+p_j}+\frac{3p_i^2k_j+3p_j^2k_i}{k_ik_j(p_i+p_j)}\right)$} & \multirow{2}{*}{$\frac{(k_i+k_j)^2}{2}$} & $u_{i+j}$ \\
&$\widehat{l_{i+j}}$ & & & $\widehat{u_{i+j}}$ \\
\hline
\multirow{2}{*}{$S_{i-j}$} & $l_{i-j}$ & \multirow{2}{*}{$\left(k_i^2+k_ik_j+k_j^2+\frac{3p_i^2k_j-3p_j^2k_i}{k_ik_j(k_i-k_j)}, \frac{k_i^3-k_j^3}{p_i-p_j}+\frac{3p_i^2k_j-3p_j^2k_i}{k_ik_j(p_i-p_j)}\right)$} & \multirow{2}{*}{$\frac{(k_i-k_j)^2}{2}$} & $u_{i-j}$ \\
&$\widehat{l_{i-j}}$ & & & $\widehat{u_{i-j}}$ \\
\hline
$S_{1+2+3}$ & $\widehat{l_{1+2+3}}$ & $\left(\frac{k_1^3+k_2^3+k_3^3}{k_1+k_2+k_3}+\frac{\delta_1}{k_1k_2k_3(k_1+k_2+k_3)}, \frac{k_1^3+k_2^3+k_3^3}{p_1+p_2+p_3}+\frac{\delta_1}{k_1k_2k_3(p_1+p_2+p_3)}\right)$ & $\frac{(k_1+k_2+k_3)^2}{2}$ & $\widehat{u_{1+2+3}}$\\
\hline
$S_{1+2-3}$ & $\widehat{l_{1+2-3}}$ & $\left(\frac{k_1^3+k_2^3-k_3^3}{k_1+k_2-k_3}+\frac{\delta_2}{k_1k_2k_3(k_1+k_2-k_3)}, \frac{k_1^3+k_2^3-k_3^3}{p_1+p_2-p_3}+\frac{\delta_2}{k_1k_2k_3(p_1+p_2-p_3)}\right)$ & $\frac{(k_1+k_2-k_3)^2}{2}$ & $\widehat{u_{1+2-3}}$\\
\hline
$S_{1-2+3}$ & $l_{1-2+3}$ & $\left(\frac{k_1^3-k_2^3+k_3^3}{k_1-k_2+k_3}+\frac{\delta_3}{k_1k_2k_3(k_1-k_2+k_3)}, \frac{k_1^3-k_2^3+k_3^3}{p_1-p_2+p_3}+\frac{\delta_3}{k_1k_2k_3(p_1-p_2+p_3)}\right)$ & $\frac{(k_1-k_2+k_3)^2}{2}$ & $u_{1-2+3}$\\
\Xhline{1pt}
\end{tabular}
\vspace{0.4\baselineskip}
\caption*{\captionsetup{justification=raggedright,singlelinecheck=false,format=hang} \quad \normalsize
In this table, $i,\,j\in\{1,2,3\}$ and 
$\delta_1=3p_1^2k_1k_2+3p_2^2k_1k_3+3p_3^2k_2k_3$, $\delta_2=3p_1^2k_1k_2+3p_2^2k_1k_3-3p_3^2k_2k_3$, $\delta_3=3p_1^2k_1k_2-3p_2^2k_1k_3+3p_3^2k_2k_3$.
The expressions of soliton arms are given by Eqs.\ \eqref{uj01},\eqref{uj02},\eqref{uj03}, and their trajectories are given by Eqs.\ \eqref{l01}, \eqref{l02}, \eqref{l03}.}
\end{table}

\section{The stem structure in 3-resonance 3-soliton}\label{sec3}
When all \(a_{ij}\) satisfy the resonance conditions (\(a_{ij}\to 0\) or \(+\infty\)), we obtain the 3-resonant 3-soliton solutions.  
Different types of such solutions arise from various combinations of resonance intensities (i.e., combinations of strong and weak resonances), as detailed below:

\textbf{Case 3.1: \(a_{12},\,a_{13},\,a_{23}\to 0\).}  
In this case, we have
\[p_1 = -\frac{k_1 (k_1 k_3 - k_3^2 + p_3)}{k_3}, \quad p_2 = -\frac{k_2 (k_2 k_3 - k_3^2 + p_3)}{k_3},\]
or
\[p_1 = \frac{k_1 (k_1 k_3 - k_3^2 + p_3)}{k_3}, \quad p_2 = \frac{k_2 (k_2 k_3 - k_3^2 + p_3)}{k_3}.\]
We consider only the former case. 
Then, the weak 3-resonant 3-soliton solution is given by
\begin{equation}\label{3f3r1}
u = 2(\ln f)_{xx}, \quad f = 1 + \exp\xi_1 + \exp\xi_2 + \exp\xi_3.
\end{equation}

\textbf{Case 3.2: $a_{12}\to 0$, $a_{13}$ and $a_{23}\to +\infty$.}  
In this case, we have
\[p_1 = -\frac{k_1 (k_1 k_3 + k_3^2 - p_3)}{k_3}, \quad p_2 = -\frac{k_2 (k_2 k_3 + k_3^2 - p_3)}{k_3},\]
or
\[p_1 = \frac{k_1 (k_1 k_3 + k_3^2 + p_3)}{k_3}, \quad p_2 = \frac{k_2 (k_2 k_3 + k_3^2 + p_3)}{k_3}.\]
We consider only the former case. 
Substituting \(\xi_1 \to \xi_1 - \ln a_{13}\), \(\xi_2 \to \xi_2 - \ln a_{23}\), and using \(a_{12}\to 0\) in Eq.~\eqref{3f}, then taking the limits \(a_{13}, a_{23} \to +\infty\), we obtain the mixed 3-resonant 3-soliton solution:
\begin{equation}\label{3f3r2}
u = 2(\ln f)_{xx}, \quad f = 1 + \exp\xi_3 + \exp(\xi_1 + \xi_3) + \exp(\xi_2 + \xi_3) .
\end{equation}

Based on the above analysis, we identify two distinct types of 3-resonant 3-soliton solutions:
the weak 3-resonant soliton (Case 3.1) and the mixed 3-resonant soliton (Case 3.2).
Among these, Case 3.1 will be the primary focus of our subsequent discussion.
Using the same asymptotic analysis method as in the previous section, we obtain the following proposition.
\begin{prop}\label{prop3.1}
The asymptotic forms of the weak 3-resonant 3-soliton given by Eq.\ \eqref{3f3r1} in case 3.1 with 
$p_1 = -\frac{k_1 ( k_1 k_3 - k_3^2 + p_3)}{k_3}, \, p_2 = -\frac{k_2 ( k_2 k_3 -k_3^2 +p_3)}{k_3}$ are as following:
\begin{flalign}\label{3s3rasy01}
\begin{split}
y\to -\infty,\quad &S_1:\quad u\sim u_1,\\
y\to +\infty,\quad &S_3:\quad u\sim u_3,\quad S_{1-2}:\quad u\sim u_{1-2},\quad S_{2-3}:\,\,u\sim u_{2-3}.
\end{split}
\end{flalign}
The stem structures:
\begin{flalign}\label{3s3rstem01}
\begin{split}
t\to-\infty,\quad &S_{1-3}:\, u\sim u_{1-3};\quad t\to+\infty,\quad S_2:\, u\sim u_2.
\end{split}
\end{flalign}
The relevant expressions are provided in Eqs.\ \eqref{uj01} and \eqref{uj02}.
\end{prop}

In contrast to the 2-resonant case, the 3-resonant soliton solution does not depend on any \( a_{ij} \), and no phase shift occurs in the soliton arms before or after the interaction. 
As a result, the asymptotic forms of the soliton arms remain identical in the limits \( t \to -\infty \) and \( t \to +\infty \).

\begin{figure}[h!tb]
\centering
\subfigure[$t=-8$]{\includegraphics[height=4cm,width=5cm]{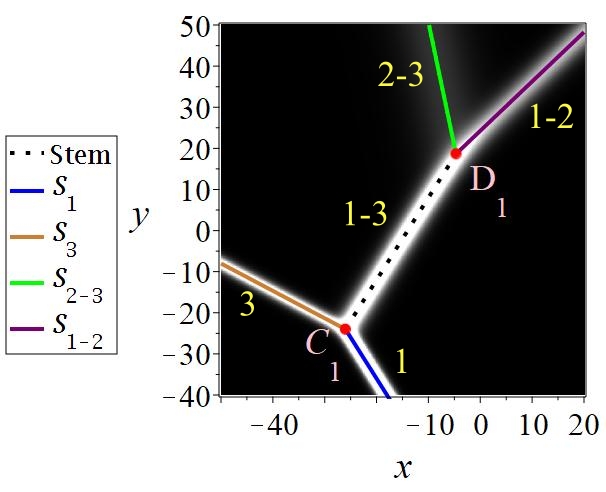}}
\subfigure[$t=0$]{\includegraphics[height=4cm,width=4cm]{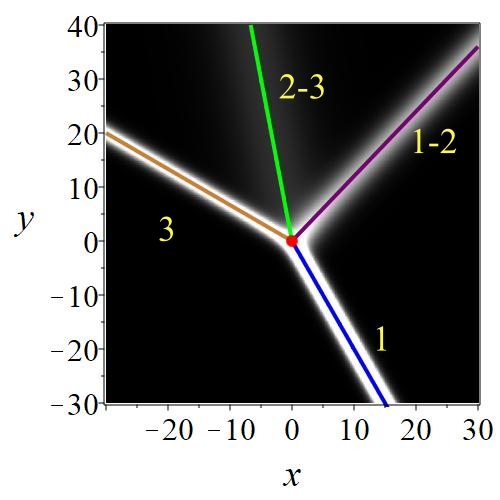}}
\subfigure[$t=10$]{\includegraphics[height=4cm,width=4cm]{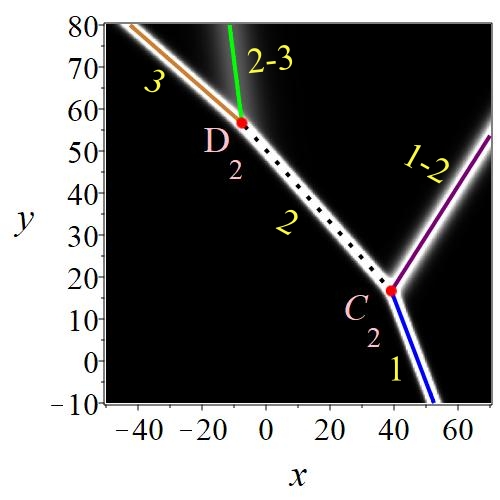}}
\vspace{-0.6\baselineskip}
\caption{The density plots of the weak 3-resonant 3-soliton with 
$k_1 = 2,\,  k_2 = \frac{4}{3},\,  k_3 = 1,\,  p_1 = -\frac{k_1 ( k_1 k_3 - k_3^2 + p_3)}{k_3}, \, p_2 = -\frac{k_2 ( k_2 k_3 -k_3^2 +p_3)}{k_3}$. 
The lines are the trajectories of the arms and stem structures, and the points are the endpoints of the variable-length stem structures.
}\label{fig3-1}
\end{figure}

The reconnection of the four soliton arms, along with the disappearance and regeneration of the stem structure, is illustrated in Fig.~\ref{fig3-1}. 
As \( t \to -\infty \), the stem structure \( S_{1-3} \) connects two pairs of V-shaped solitons: \( (S_{1-2},\, S_{2-3}) \) and \( (S_1,\, S_3) \). 
As time evolves, the length of this stem gradually decreases and vanishes around \( t = 0 \). 
At this moment, the four arms---\( S_1,\, S_3,\, S_{1-2},\, S_{2-3} \)---come together and join at a common location, and the two original V-shaped pairs transform into new ones: \( (S_1,\, S_{1-2}) \) and \( (S_3,\, S_{2-3}) \). 
As time further increases (\( t \to +\infty \)), a new stem structure \( S_2 \) emerges, gradually elongating to connect the two new V-shaped soliton pairs.

The trajectories, amplitudes, and velocities of the four arms and two stems are summarized in Table~\ref{tab:t1}. 
By computing the intersection points of these trajectories, the endpoints of the variable-length stem structures can be explicitly determined as follows:
\begin{flalign}\label{endpoints4}
\begin{split}
C_1\,&\Bigg(\bigg(k_3^2-2p_3+\frac{4k_1p_3}{k_3}-\frac{3p_3^2}{k_3^2}\bigg)t,\,\bigg(\frac{6p_3}{k_3}-4k_1+2k_3\bigg)t \Bigg),\\
D_1\,&\Bigg(\bigg(k_3^2-4k_1k_2-2p_3-\frac{4k_1p_3+4p_3k_2}{k_3}-\frac{3p_3^2}{k_3^2}\bigg)t,\,\bigg(\frac{6p_3}{k_3}-4k_1-4k_2+2k_3\bigg)t \Bigg),\\
C_2\,&\Bigg(\bigg(-3k_3^2+4k_1k_3+4k_2k_3-4k_1k_2-6p_3+\frac{4k_1p_3+4p_3k_2}{k_3}-\frac{3p_3^2}{k_3^2}\bigg)t,\,\bigg(\frac{6p_3}{k_3}-4k_1-4k_2+6k_3\bigg)t \Bigg),\\
D_2\,&\Bigg(\bigg(k_3^2-2p_3-\frac{4p_3k_2}{k_3}-\frac{3p_3^2}{k_3^2}\bigg)t,\,\bigg(\frac{6p_3}{k_3}-4k_2+2k_3\bigg)t \Bigg).
\end{split}
\end{flalign}
In this setting, point \( C_1 \) represents the intersection of lines \( l_1 \) and \( l_3 \), while point \( D_1 \) corresponds to the intersection of \( l_{1-2} \) and \( l_{2-3} \).  
Similarly, point \( C_2 \) is defined by the intersection of \( l_1 \) and \( l_{1-2} \), and point \( D_2 \) by that of \( l_3 \) and \( l_{2-3} \).  
Unlike the 2-resonant case, the two endpoints of the stem structure coincide at \((0, 0)\) when \( t = 0 \), indicating that the switch between \( S_{1-3} \) and \( S_2 \) occurs precisely at this moment.
These characteristic points are illustrated in Fig.~\ref{fig3-1}.  
Based on this, the lengths of the variable-length stem trajectories can be determined as follows:
\begin{flalign}\label{3length-4}
\begin{split}
&|C_1D_1|=4|k_2t|\sqrt{k_1^2+1-\frac{2k_1p_3}{k_3}+\frac{p_3^2}{k_3^2}},\,t\leqslant 0,\\
&|C_2D_2|=4|t(k_1-k_3)|\sqrt{(k_2-k_3)^2+2p_3+1-\frac{2k_2p_3}{k_3}+\frac{p_3^2}{k_3^2}},\,t> 0.
\end{split}
\end{flalign}

Next, we analyze the amplitudes of the variable-length stem structures.  
The cross-sectional profiles of the 3-soliton solution \eqref{3f}, given by Eqs.~\eqref{3f3r1} along the lines \( l_{1-3} \) and \( l_2 \) shown in Fig.~\ref{fig3-1}, are expressed as follows:
\begin{flalign}\label{cross3s04}
\begin{split}
u|_{l_{1-3}}=\frac{4(k_1-k_2)^2\e^{\theta_{13}+\theta_{14}}+4k_1^2\e^{\theta_{13}}+4k_2^2\e^{\theta_{14}}}{(1+2\e^{\theta_{13}}+\e^{\theta_{14}})^2},\,
u|_{l_2}=\frac{2(k_1-k_3)^2\e^{\theta_{15}+\theta_{16}}+4k_1^2\e^{\theta_{15}}+4k_3^2\e^{\theta_{16}}}{(2+\e^{\theta_{15}}+\e^{\theta_{16}})^2},
\end{split}
\end{flalign}
where
\begin{flalign*}
&\theta_{13}=\frac{(k_1p_3-k_3p_1)y}{k_1-k_3}+\bigg(k_1k_2(k_1+k_3)-\frac{3k_1^2p_3^2-3k_3^2p_1^2}{k_1k_3(k_1-k_3)}\bigg)t,\\
&\theta_{14}=\bigg(p_2-\frac{k_2p_1-k_2p_3}{k_1-k_3}\bigg)y+\bigg(k_2(k_1^2+k_1k_3+k_3^2)-k_3^2-\frac{3p_2^3}{k_2}
-\frac{3k_1k_2p_3^2-3p_1k_2k_3}{k_1k_3(k_1-k_3)}\bigg)t,\\
&\theta_{15}=\bigg(p_1-\frac{k_1p_2}{k_2}\bigg)y-\bigg(k_1^3-k_1k_2^2+\frac{3p_1^2}{k_1}-\frac{3k_1p_2^2}{k_2^2}\bigg)t,\\
&\theta_{16}=\bigg(p_3-\frac{k_3p_2}{k_2}\bigg)y-\bigg(k_3^3-k_2^2k_3+\frac{3p_3^2}{k_3}-\frac{3k_3p_2^2}{k_2^2}\bigg)t.
\end{flalign*}

\begin{figure}[h!tb]
\centering
\subfigure[$t=-8$]{\includegraphics[height=3cm,width=3cm]{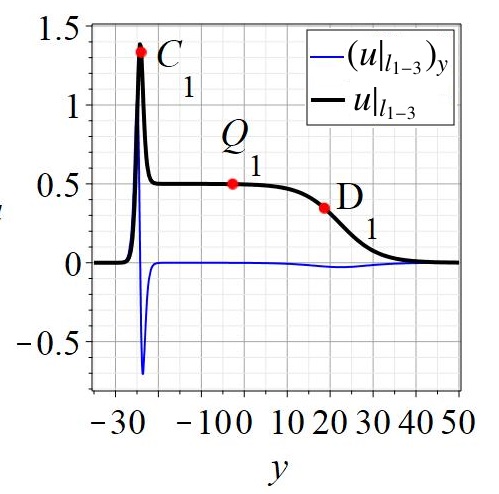}}
\subfigure[$t=10$]{\includegraphics[height=3cm,width=3cm]{3s3weak-15.jpg}}
\subfigure[]{\includegraphics[height=3cm,width=3cm]{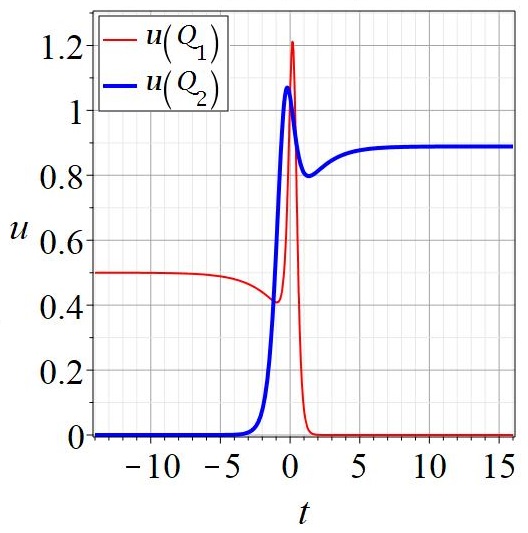}}
\subfigure[$t=-8$]{\includegraphics[height=3cm,width=3cm]{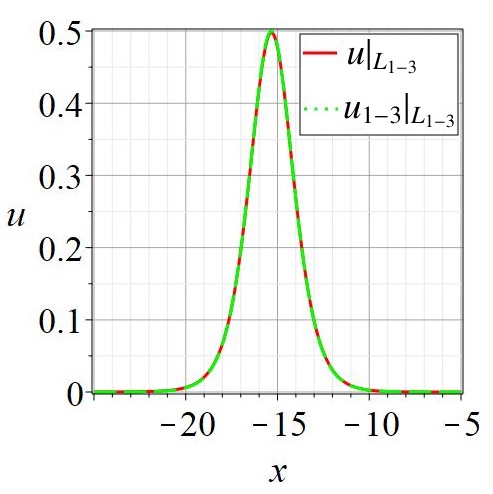}}
\subfigure[$t=10$]{\includegraphics[height=3cm,width=3cm]{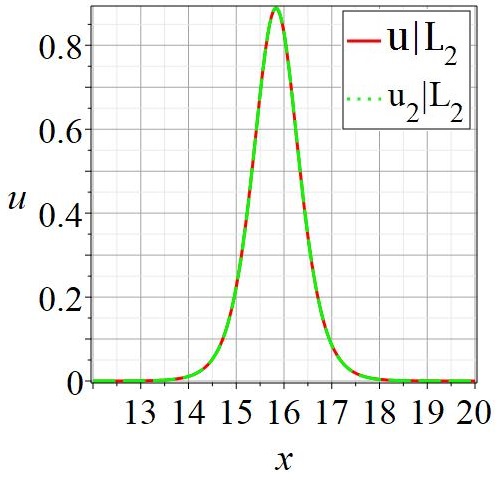}}
\vspace{-0.6\baselineskip}
\caption{The density plots of the weak 3-resonant 3-soliton with 
$k_1 = 2,\,  k_2 = \frac{4}{3},\,  k_3 = 1,\,  p_1 = -\frac{k_1 ( k_1 k_3 - k_3^2 + p_3)}{k_3}, \, p_2 = -\frac{k_2 ( k_2 k_3 -k_3^2 +p_3)}{k_3}$. 
The lines are the trajectories of the arms and stem structures, and the points are the endpoints of the variable length stem structures.
}\label{fig3-2}
\end{figure}

The graphs of $u|_{l_{1-3}}$ and $(u|_{l_{1-3}})_y$ are presented in Fig.\ \ref{fig3-2} (a). 
Unlike the scenario in Section 2, it is not possible to derive the amplitude of $u|_{l_{1-3}}$ by finding the extreme points of $u|_{l_{1-3}}$ using \((u|_{l_{1-3}})_y = 0\) in this case. 
From the graphs of $u|_{l_{1-3}}$ and $(u|_{l_{1-3}})_y$, we observe that the amplitude varies minimally near the midpoint of the stem (where \((u|_{l_{1-3}})_y \approx 0\)). 
Therefore, we focus our study on the amplitude at the midpoint of $C_1D_1$ and $C_2D_2$, which are denoted as $Q_1$ and $Q_2$, are expressed as:
\begin{flalign} 
\begin{split}\label{up04}
u(Q_1)=\frac{2\e^{-\frac{128t}{27}} \left(9\e^{-\frac{160t}{27}} + 5\e^{-\frac{16t}{3}} + 45\e^{-\frac{16t}{27}} + 16\right)}{9\left(1 + 2\e^{-\frac{16t}{3}} + \e^{-\frac{128t}{27}}\right)^2},\,
u(Q_2)=\frac{2\left(9\e^{-\frac{10t}{3}} + 40\e^{-\frac{8t}{3}} + 10\e^{-\frac{2t}{3}} + 16\right)}{9\left(2 + \e^{-\frac{8t}{3}} + \e^{-\frac{2t}{3}}\right)^2}.
\end{split}
\end{flalign}
The amplitude trend plots of $u(Q_1)$ and $u(Q_2)$ are shown in Fig.\ \ref{fig3-2} (a). 
As can be seen from the figure, the amplitude changes violently and complex near $t=0$, and gradually flattens out as t approaches infinity. 
This is because near $t=0$, the four arms are close together and interact strongly, resulting in the presence of the stem not being obvious. 
Studying the properties of stems at this point is of little significance. 
Due to $\lim\limits_{t\to -\infty}u(Q_1)=\frac{1}{2}=\frac{(k_1-k_3)^2}{2}$, $\lim\limits_{t\to +\infty} u(Q_2)=\frac{8}{9}=\frac{k_2^2}{2}$, 
it is reliable that we use $u(Q_1)$ and $u(Q_2)$ as approximations of amplitude of the stem structures $S_{1-3}$ and $S_2$.

In addition, we consider the vertical plane passing through points $Q_1$ and perpendicular to $l_{1-3}$ is
$$L_{1-3}^{(2)}:\quad (p_1-p_3)x-(k_1-k_3)y-[(p_1-p_3)x_{Q_1}-(k_1-k_3)y_{Q_1}]=0,$$
where $x_{Q_1}$ and $y_{Q_1}$ are the are the $(x,\,y)$ coordinates of the point $Q_1$, respectively. 
Fig.\ \ref{fig3-2} (b) shows the intersection curves of $u$ and $u_{1-3}$ with $L_{1-3}^{(2)}$, 
which denoted as $u|_{L_{1-3}^{(2)}}$ and $u_{1-3}|_{L_{1-3}^{(2)}}$, respectively. 

The vertical plane passing through points $Q_2$ and perpendicular to $l_2$ as
$$L_2:\quad  p_2x-k_2y-(p_2x_{Q_2}-k_2y_{Q_2})=0,$$
where $x_{Q_2}$ and $y_{Q_2}$ are the $(x,\,y)$ coordinates of the point $Q_2$, respectively. 
Fig.\ \ref{fig3-2} (c) shows the intersection curves of $u$ and $u_2$ with $L_2$, 
which denoted as $u|_{L_2}$ and $u_2|_{L_2}$, respectively. 
It can be seen from Fig.\ \ref{fig3-2} (c) and (d) that they almost coincide. 
This also shows that it is feasible to use $u_{1-3}$ and $u_2$ as the approximation of $u$ for $t\to-\infty$ and $t\to+\infty$.

\begin{figure}[h!tb]
\centering
\subfigure[$t=-8$]{\includegraphics[height=4cm,width=5cm]{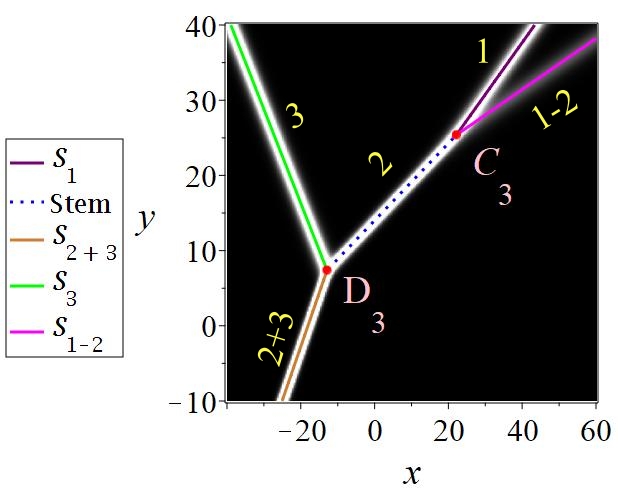}}
\subfigure[$t=0$]{\includegraphics[height=4cm,width=4cm]{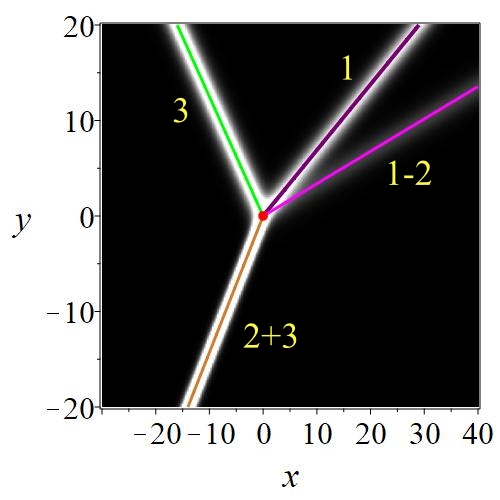}}
\subfigure[$t=10$]{\includegraphics[height=4cm,width=4cm]{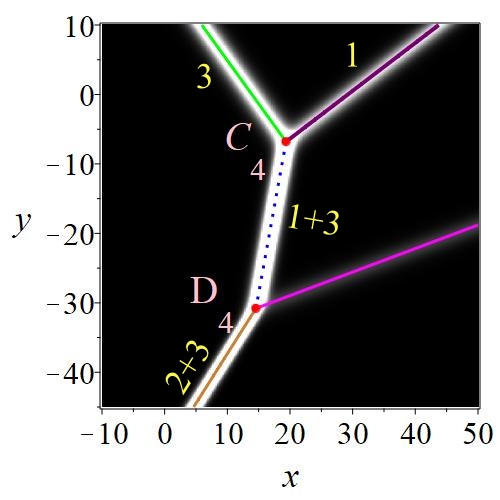}}
\vspace{-0.6\baselineskip}
\caption{parameters: $k_1=2,\,k_2=-1,\,k_3=-\frac{3}{2},\,p_1 = \frac{k_1 \left( k_1 k_3 + k_3^2 + p_3 \right)}{k_3}, \, p_2 = \frac{k_2 \left( k_2 k_3 -k_3^2 +p_3 \right)}{k_3},\,p_3=1$. 
(a) The cross-sectional curves $u|_{l_{1-3}}$ given by \eqref{cross3s04}; 
(b) The cross-sectional curves $u|_{l_2}$ given by \eqref{cross3s04}; 
(c) The amplitude evolution curves \eqref{up04}; 
(d) The cross-sectional curves $u|_{L_{2-3}^{(2)}}$ and $\widehat{u_{2-3}}|_{L_{2-3}^{(2)}}$; 
(e) The cross-sectional curves $u|_{L_{1+3}}$ and $u_{1+3}|_{L_{1+3}}$.
}\label{fig3-3}
\end{figure}

The analysis of the weak 3-resonant 3-soliton in Case 3.1 has been completed. 
Since the method of analyzing the mixed 3-resonant 3-soliton in Case 3.2 is similar, we will omit a detailed discussion. 
Fig.~\ref{fig3-3} illustrates the asymptotic form, trajectory, and stem structure endpoint. 
The corresponding formulas have already been provided above.
Only the asymptotic form of the mixed 3-resonant 3-soliton \eqref{3f3r2} is presented below:
\begin{prop}\label{prop3.2}
The asymptotic form of the mixed 3-resonant 3-soliton \eqref{3f3r2} is 
\begin{flalign}\label{3s3rasy02}
\begin{split}
y\to -\infty,\quad &S_{2+3}:\quad u\sim u_{2+3},\\
y\to +\infty,\quad &S_1:\quad u\sim u_1,\quad S_3:\quad u\sim u_3,\quad S_{1-2}:\quad u\sim u_{1-2}.
\end{split}
\end{flalign}
The stem structures:
\begin{flalign}\label{3s3rstem02}
\begin{split}
t\to-\infty,\,&S_2:\, u\sim u_2;\quad t\to+\infty,\,S_{1+3}:\, u\sim u_{1+3}.
\end{split}
\end{flalign}
\end{prop}

\section{Conclusions and discussions}\label{summary}
This paper provides a systematic study of the asymptotic forms and variable-length stem structures in 2-resonant and 3-resonant 3-soliton solutions of the KPII equation.  
In both types of resonant soliton interactions, a phenomenon of soliton reconnection occurs: 
the ends of a variable-length stem structure connect to the vertices of two V-shaped solitons, and the soliton arms undergo reconnection, while the stem structure vanishes and reappears across the interaction.  
Each type of resonant soliton can be further classified into \textbf{strong}, \textbf{weak}, or \textbf{mixed (strong--weak)} resonance, according to the specific resonance conditions imposed on the limit of phase shift parameters $a_{ij}$, following the classification scheme of Ref.~\cite{jpsj1983}.

For the \textbf{2-resonant 3-soliton solution}, we categorized three distinct cases---strong, weak, and mixed 2-resonances---whose asymptotic forms are given in Propositions~\ref{prop2.1}--\ref{prop2.3}. 
Based on these asymptotic forms and the corresponding stem structures, we highlight the following observations:

\begin{itemize}[itemsep=0.5em, parsep=0pt, topsep=0.3em, leftmargin=2em]
\item The parameter \(a_{12}\) induces distinct asymptotic configurations before and after interaction, leading to a phase shift \(\ln a_{12}\) between soliton arms \(S_1\) and \(S_2\). 
In the strong 2-resonant 3-soliton case, we obtain the correct asymptotic form of the stem structure and its soliton arms, which substantially extends and corrects the description in Ref.~\cite{jpsj1983}, where only the near \(t=0\) configuration was implicitly captured.

\item Analytical expressions for the endpoints, lengths, and extrema of stem structures are derived, yielding a complete analytical characterization of their spatial localization in the asymptotic regimes \(t\to\pm\infty\). 
The stem extreme line is shown to be a curved trajectory rather than a straight one, with a nearly constant amplitude around its midpoint, revealing a nontrivial geometric feature absent in standard line-soliton theory.

\item The reconnection of stem structures is rigorously identified as a limiting process associated with \(a_{13}, a_{23}\to 0\) or \(\infty\), providing an analytical explanation of the topological transition observed in resonant 3-soliton interactions, which corresponds to Figs. \ref{fig2-1}--\ref{fig2-7} and \ref{fig2-9}.
\end{itemize}

For the \textbf{3-resonant 3-soliton case}, we presented two resonant soliton solutions (Cases 3.1--3.2), associated with different combinations of resonance conditions, along with their corresponding asymptotic forms.  
Focusing on Case 3.1 (the weak 3-resonant case), we applied a similar methodology to analyze its local properties---such as soliton trajectories, stem endpoints and lengths, cross---sectional profiles, and amplitude variations. 
Compared with the 2-resonant case, several key differences emerge:

\begin{itemize}[itemsep=0.5em, parsep=0pt, topsep=0.3em, leftmargin=2em]
\item For the 3-resonant soliton, no phase shift occurs between soliton arms, and the interaction manifests solely through the creation and annihilation of the stem structure. The four surrounding soliton arms remain asymptotically invariant as \(t\to\pm\infty\).

\item The stem structure disappears and re-emerges exactly at \(t=0\), where all four soliton arms intersect at a single point. This yields a globally valid analytical formula for the stem length, in contrast to the 2-resonant case.

\item The reconnection of stem structures is obtained as a well-defined limiting process \(a_{ij}\to 0\) or \(\infty\), giving a rigorous analytical description of the associated topological transition in 3-resonant soliton interactions, which corresponds to Figs. \ref{fig3-2} and \ref{fig3-3}.

\end{itemize}

It is particularly noteworthy that both this work and Ref.~\cite{kp2025} investigate the stem structures of solitons in the KPII equation; 
however, the research objects and the resulting conclusions differ substantially:

\begin{itemize}
\item Quasi-resonant 2-soliton solutions (Ref.~\cite{kp2025}):  
These correspond to the case \(a_{12} \approx 0\) or \(\infty\).  
In this situation, the shape of the quasi-resonant 2-soliton solution remains invariant during time evolution, and consequently, the associated stem structures also do not change with time.  
Their lengths are independent of \(t\), and the asymptotic form of the soliton does not require consideration of the limits \(t \to \pm \infty\).
Note that in this case, the soliton does not exhibit the reconnection phenomenon.

\item Resonant 3-soliton solutions (this work):  
These correspond to the case \(a_{ij} \to 0\) or \(\infty\).  
In this setting, the soliton shape varies with time, so the asymptotic forms must be analyzed in the limits \(t \to \pm \infty\).  
Moreover, during the evolution process, existing stem structures annihilate while new ones emerge, and the length of the stem structures changes dynamically with time.
In particular, the 3-soliton undergoes reconnection in this case.
\end{itemize}

A future research direction is to investigate such localized structures within high-order stem structures of the KPI equation.

\section*{Appendix}\addcontentsline{toc}{section}{Appendix}

\textbf{Case 2.2:} Taking parameter $k_1=-\frac{2}{3},\,k_2=-1,\,k_3=\frac{4}{3},\,p_1 = \frac{k_1 \left( k_1 k_3 + k_3^2 + p_3 \right)}{k_3}, \, p_2 = -\frac{k_2 \left( k_2 k_3 + k_3^2 -p_3 \right)}{k_3},\,p_3=\frac{2}{3}$, 
we obtain the asymptotic of \eqref{3f2r2} as follows:

Before collision ($t\to-\infty$):
\begin{flalign}
\begin{split}
y\to -\infty,\quad&S_1:\quad  u\sim u_1,\quad S_2:\quad u\sim \widehat{u_2}, \quad S_3:\quad u\sim u_3;\\
y\to +\infty,\quad&S_{1+2+3}:\,\,u\sim \widehat{u_{1+2+3}}.
\end{split}
\end{flalign}

After collision ($t\to+\infty$):
\begin{flalign}
\begin{split}
y\to -\infty,\quad&S_1:\quad u\sim \widehat{u_1},\quad S_2:\quad u\sim u_2, \quad S_3:\quad u\sim u_3;\\
y\to +\infty,\quad&S_{1+2+3}:\,\,u\sim \widehat{u_{1+2+3}}.
\end{split}
\end{flalign}

The stem structures:
\begin{flalign}
\begin{split}
t\to-\infty,\quad&S_{2+3}:\quad u\sim \widehat{u_{2+3}};\quad t\to+\infty,\,S_{1+3}:\, u\sim \widehat{u_{1+3}}.
\end{split}
\end{flalign}

\begin{figure}[h!tb]
\centering
\subfigure[$t=-1$]{\includegraphics[height=4cm,width=5.5cm]{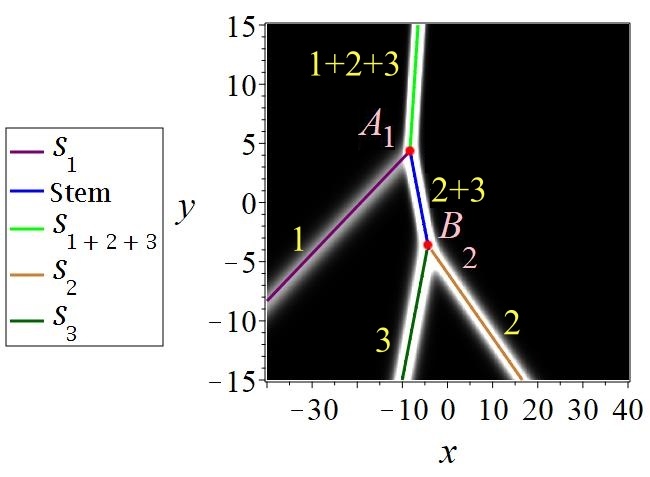}}
\subfigure[$t=0$]{\includegraphics[height=4cm,width=4cm]{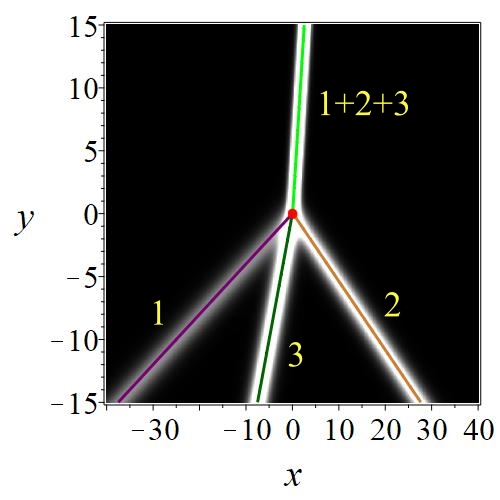}}
\subfigure[$t=2$]{\includegraphics[height=4cm,width=4cm]{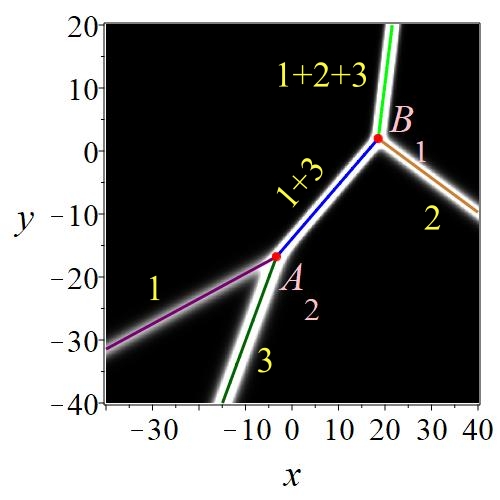}}
\vspace{-0.6\baselineskip}
\caption{The density plots of the strong 2-resonance 3-soliton \eqref{3f2r2} with 
$k_1=-\frac{2}{3},\,k_2=-1,\,k_3=\frac{4}{3},\,p_1 = \frac{k_1 \left( k_1 k_3 + k_3^2 + p_3 \right)}{k_3}, \, p_2 = -\frac{k_2 \left( k_2 k_3 + k_3^2 -p_3 \right)}{k_3},\,p_3=\frac{2}{3}$. 
The lines are the trajectories of the arms and stem structures, and the points are the endpoints of the variable length stem structures.
}\label{fig2-4}
\end{figure}

By solving for the intersection points of these trajectories, the endpoints of the variable-length stem structures are $A_1,\,A_2,\,B_1,\,B_2$ 
which have given by \eqref{endpoints1}. 
The density plots of their temporal evolution and the soliton arm trajectories are presented in Fig.\ \ref{fig2-4}

\textbf{Case 2.3:} Taking parameters 
$k_1=-1,\,k_2=-\frac{2}{3},\,k_3=-\frac{4}{3},\,p_1 = \frac{k_1 \left( k_1 k_3 + k_3^2 + p_3 \right)}{k_3}, \, p_2 = -\frac{k_2 \left( k_2 k_3 + k_3^2 -p_3 \right)}{k_3},\,p_3=\frac{2}{3}$, 
we obtain the asymptotic of \eqref{3f2r3} as follows:

Before collision ($t\to-\infty$):
\begin{flalign}
\begin{split}
y\to -\infty,\quad&S_{1+2+3}:\,\,u\sim \widehat{u_{1+2+3}};\\
y\to +\infty,\quad&S_1:\quad  u\sim u_1,\quad S_2:\quad u\sim \widehat{u_2}, \quad S_3:\quad u\sim u_3.
\end{split}
\end{flalign}

After collision ($t\to+\infty$):
\begin{flalign}
\begin{split}
y\to -\infty,\quad&S_{1+2+3}:\,\,u\sim \widehat{u_{1+2+3}};\\
y\to +\infty,\quad&S_1:\quad u\sim \widehat{u_1},\quad S_2:\quad u\sim u_2, \quad S_3:\quad u\sim u_3.
\end{split}
\end{flalign}

The stem structures:
\begin{flalign}
\begin{split}
t\to-\infty,\quad&S_{1+3}:\quad u\sim u_{1+3};\quad t\to+\infty,\,S_{2+3}:\, u\sim u_{2+3}.
\end{split}
\end{flalign}

\begin{figure}[h!tb]
\centering
\subfigure[$t=-1$]{\includegraphics[height=4cm,width=5.5cm]{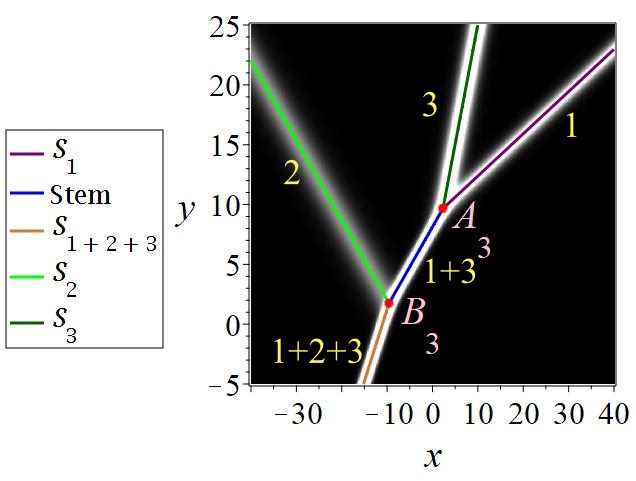}}
\subfigure[$t=0$]{\includegraphics[height=4cm,width=4cm]{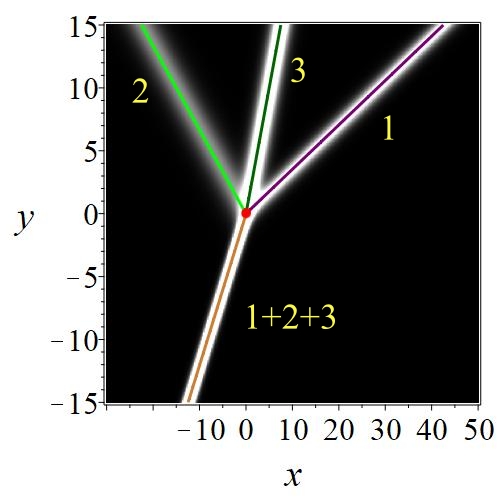}}
\subfigure[$t=1$]{\includegraphics[height=4cm,width=4cm]{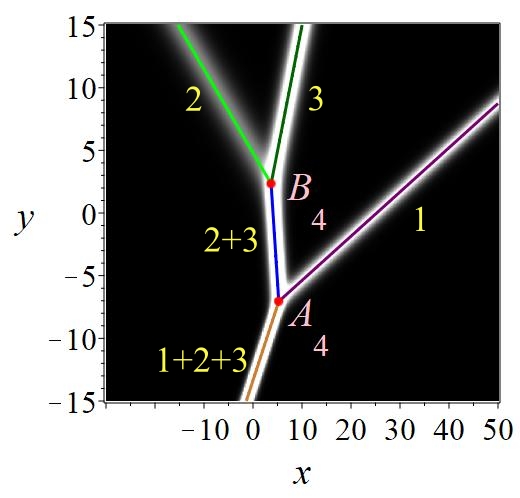}}
\vspace{-0.6\baselineskip}
\caption{The density plots of the strong 2-resonance 3-soliton \eqref{3f2r3} with 
$k_1=-1,\,k_2=-\frac{2}{3},\,k_3=-\frac{4}{3},\,p_1 = \frac{k_1 \left( k_1 k_3 + k_3^2 + p_3 \right)}{k_3}, \, p_2 = -\frac{k_2 \left( k_2 k_3 + k_3^2 -p_3 \right)}{k_3},\,p_3=\frac{2}{3}$. 
The lines are the trajectories of the arms and stem structures, and the points are the endpoints of the variable length stem structures.
}\label{fig2-5}
\end{figure}

By solving for the intersection points of these trajectories, one can ascertain the endpoints of the variable-length stem structures as follows,
\begin{flalign}\label{endpoints5}
\begin{split}
A_3\,&\Bigg(-\bigg(\frac{4k_1k_3p_3+3p_3^2}{k_3^2}-k_3^2+2p_3\bigg)t,\,\bigg(\frac{6p_3}{k_3}+4k_1+2k_3\bigg)t \Bigg),\\
B_3\,&\Bigg(-\frac{(k_1k_3+p_3)\ln a_{12}}{k_2k_3(k_1+k_2+k_3)}+\bigg(k_3^2+4k_1k_2+4k_1k_3+2p_3+\frac{4k_2p_3-4p_3k_1}{k_3}-\frac{3p_3^2}{k_3^2}\bigg)t,\\
&\frac{\ln a_{12}}{k_2(k_1+k_2+k_3)}+\bigg(\frac{6p_3}{k_3}+4k_1-4k_2-2k_3\bigg)t\Bigg),\\
A_4\,&\Bigg(-\frac{(k_2k_3-p_3)\ln a_{12}}{k_1k_3(k_1+k_2+k_3)}+\bigg(k_3^2+4k_1k_2+4k_2k_3-2p_3-\frac{4k_1p_3-4p_3k_2}{k_3}-\frac{3p_3^2}{k_3^2}\bigg)t,\\
&-\frac{\ln a_{12}}{k_1(k_1+k_2+k_3)}+\bigg(\frac{6p_3}{k_3}+4k_1-4k_2+2k_3\bigg)t\Bigg),\\
B_4\,&\Bigg(\bigg(\frac{4k_2k_3p_3-3p_3^2}{k_3^2}+k_3^2+2p_3\bigg)t,\,\bigg(\frac{6p_3}{k_3}-4k_2-2k_3\bigg)t \Bigg).
\end{split}
\end{flalign}
The density plots of their temporal evolution and the soliton arm trajectories are presented in Fig.\ \ref{fig2-5}
\textbf{Case 2.4:} Taking parameters 
$k_1=2,\,k_2=1,\,k_3=\frac{2}{3},\,p_1 = \frac{k_1 \left( k_1 k_3 + k_3^2 + p_3 \right)}{k_3}, \, p_2 = -\frac{k_2 \left( k_2 k_3 + k_3^2 -p_3 \right)}{k_3},\,p_3=\frac{3}{2}$, 
we obtain the asymptotic of \eqref{3f2r4} as follows:

Before collision ($t\to-\infty$):
\begin{flalign}
\begin{split}
x\to -\infty,\,&S_2:\quad u\sim \widehat{u_2},\quad S_{1+3}:\,\,u\sim u_{1+3},\\
x\to +\infty,\,&S_1:\quad u\sim \widehat{u_1}\quad S_{2+3}:\,\,u\sim u_{2+3}.
\end{split}
\end{flalign}

After collision ($t\to+\infty$):
\begin{flalign}
\begin{split}
x\to -\infty,\,&S_2:\quad u\sim u_2,\quad S_{1+3}:\quad u\sim u_{1+3},\\
x\to +\infty,\,&S_1:\,\,u\sim u_1,\quad S_{2+3}:\,\,u\sim u_{2+3}.
\end{split}
\end{flalign}

The stem structures:
\begin{equation}
t\to-\infty,\quad S_{1+2+3}:\, u\sim \widehat{u_{1+2+3}};\qquad t\to+\infty,\quad S_3:\,u\sim u_3.
\end{equation}

\begin{figure}[h!tb]
\centering
\subfigure[$t=-2$]{\includegraphics[height=4cm,width=5cm]{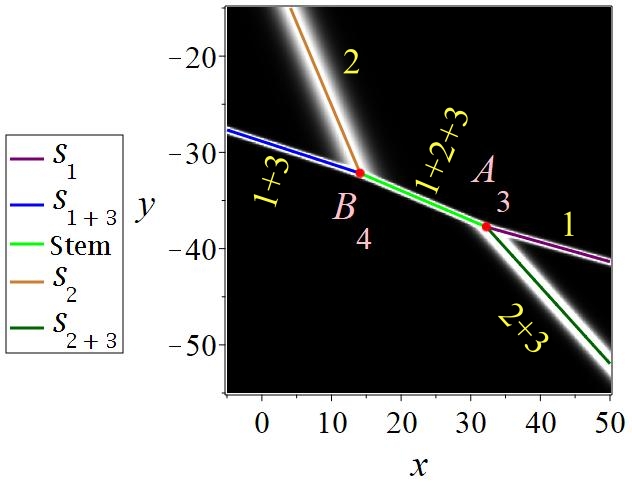}}
\subfigure[$t=0$]{\includegraphics[height=4cm,width=4cm]{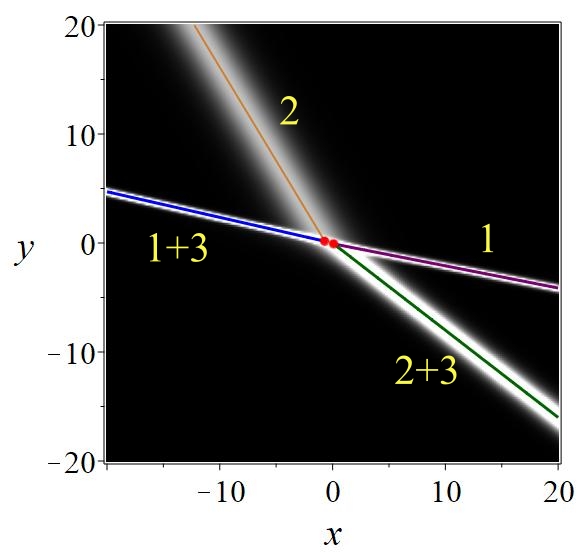}}
\subfigure[$t=1$]{\includegraphics[height=4cm,width=4cm]{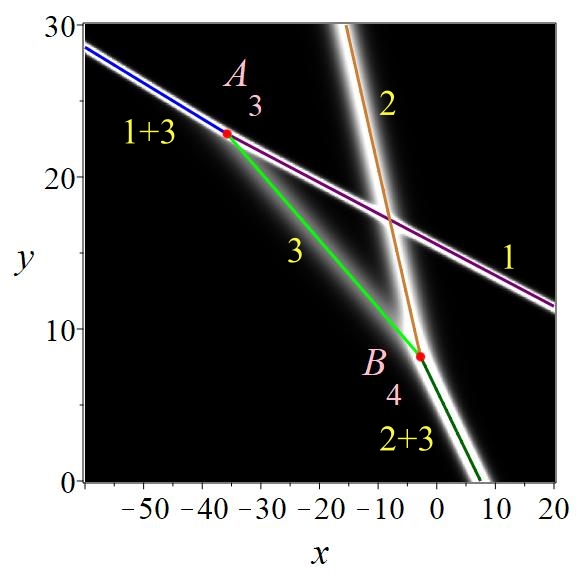}}
\vspace{-0.6\baselineskip}
\caption{The density plots of the strong 2-resonance 3-soliton \eqref{3f2r4} with 
$k_1=2,\,k_2=1,\,k_3=\frac{2}{3},\,p_1 = \frac{k_1 \left( k_1 k_3 + k_3^2 + p_3 \right)}{k_3}, \, p_2 = -\frac{k_2 \left( k_2 k_3 + k_3^2 -p_3 \right)}{k_3},\,p_3=\frac{3}{2}$. 
The lines are the trajectories of the arms and stem structures, and the points are the endpoints of the variable length stem structures.
}\label{fig2-6}
\end{figure}

By solving for the intersection points of these trajectories, the endpoints of the variable-length stem structures are $A_3,\,A_4,\,B_3,\,B_4$ 
which have given by \eqref{endpoints2}. 
The density plots of their temporal evolution and the soliton arm trajectories are presented in Fig.\ \ref{fig2-6}

\section*{Acknowledgements and declarations}
\noindent\textbf{Conflict statement}
There is neither Conflict of interest nor additional data available for this article.

\noindent\textbf{Data availability}
The data that support the findings of this study are available within the article.

\noindent\textbf{Acknowledgments}
This work is supported by the National Natural Science Foundation of China (Grant 12471239), 
and Guangdong Basic and Applied Basic Research Foundation (Grant 2024A1515013106).


\end{document}